\documentclass[british,english]{ws-rv975x65}
\usepackage{subfigure}     
\usepackage{ws-rv-van}     

\usepackage[T1]{fontenc}
\usepackage[latin9]{inputenc}
\usepackage{amsmath}
\usepackage{amssymb}
\usepackage{graphicx}
\usepackage{esint}
\usepackage{bm}
\usepackage{babel}
\usepackage[hidelinks]{hyperref}

\newcommand{\lyxdot}{.}

\makeindex

\begin{document}

\chapter[Early thermalization, hydrodynamics and energy loss in AdS/CFT]{Early thermalization, hydrodynamics and energy loss in AdS/CFT}

\author{Paul~M.~Chesler}
\vspace{-0.4cm}
\begin{center}
\textit{Department of Physics, Harvard University, Cambridge, MA 02138, USA}
\end{center}

\author[Paul~M.~Chesler and Wilke van der Schee]{Wilke van der Schee}
\vspace{-0.4cm}

\begin{center}
\textit{MIT Center for Theoretical Physics, Cambridge, MA 02139, USA}
\end{center}



\begin{abstract}
Gauge/gravity duality has provided unprecedented opportunities
to study dynamics in certain strongly coupled gauge theories.
This review aims to highlight several applications
to heavy ion collisions including far-from-equilibrium dynamics, 
hydrodynamics and jet energy loss at strong coupling.
\end{abstract}

\body
\section{Introduction}

Gauge/gravity duality equates certain gauge theories with theories of quantum gravity in one higher dimension \cite{Maldacena:1997zz}. 
This extra dimension
has a natural interpretation as the renormalization group scale, and
has led to the name ``holographic duality."
In the limit where the gauge theory is strongly coupled and has a large number of colors $N_c$
the dual description reduces to classical supergravity.  Hence, challenging strongly coupled 
quantum dynamics in the gauge theory can be accessed by solving classical 
partial differential equations.  All physics --- from microscopic interactions to macroscopic hydrodynamics --- is encoded in the 
dual classical dynamics.  Holography can thereby provide systematic and controlled access to strongly coupled dynamics via the solution
to the classical partial differential equations.
Currently there is no other theoretical tool capable of accessing strongly coupled real-time dynamics in 
in a controlled setting.  

Heavy ion collisions at RHIC and the LHC have demonstrated that the produced quark-gluon plasma is strongly coupled \cite{Aamodt:2010pa}. Hence it is of interest to model heavy ion collisions using holography. However, while there exists many theories with dual gravitational descriptions --- some with QCD-like features such as confinement and chiral symmetry breaking \cite{Witten:1998zw,Karch:2002sh} --- 
the dual description of QCD is not known (if it exists).  Nevertheless, given the immense challenge of studying strongly coupled dynamics 
in QCD it is invaluable to have a model where strongly coupled dynamics can be studied in a controlled systematic setting. This is especially true if the results are valid for a class of strongly coupled gauge theories and hence have some degree of universality.

In this review we shall focus on the simplest theory with a holographic dual: 
$3+1$ dimensional super-Yang-Mills theory with four supersymmetries ($\mathcal{N}=4$  SYM). In the limit where the SYM is strongly coupled and has a large number of colors $N_c$
the dual description reduces to classical supergravity in asymptotically five dimensional Anti-de Sitter spacetime AdS$_5$ \cite{Maldacena:1997zz} (see also \cite{Witten:1998qj,Aharony:1999ti}).
While the ground state of SYM is very different from QCD --- SYM is conformal and contains 
no particles whereas QCD is confining and has a rich spectrum of hadrons --- at temperatures not too far above the deconfinement 
transition both theories consist of strongly coupled non-Abelian plasmas.  It is in this setting where dynamics in 
SYM can potentially provide lessons for dynamics in QCD.

In what follows we highlight insights obtained via holography in three settings.  First, 
we discuss relativistic hydrodynamics including 
computations of the shear viscosity at strong coupling and the construction 
of causal viscous hydrodynamics.  We then discuss far-from-equilibrium dynamics 
and the collisions of shock waves in SYM and their application to early time dynamics 
in heavy-ion collisions.  Finally, we give a brief review of progress made in jet quenching 
and energy loss at strong coupling.

\section{First successes: viscosity and relativistic hydrodynamics}

\subsection{Viscosity from black hole horizons}

Gauge/gravity duality provides a very natural way to describe
thermal states: they are described by black holes, which are the simplest
stationary states in $AdS_{5}$ gravity with a notion of temperature.
This temperature is simply the black hole's Hawking temperature \cite{Hawking:1974sw}.
The topology of the horizon is
inherited from the gauge theory, where a thermal state usually is
extended infinitely; this is one major difference from ordinary black
holes in our universe, and in fact these black holes in $AdS_{5}$
are more appropriately called black branes. Another major difference
is the boundary of the $AdS_{5}$ spacetime, which can be thought
of as the place where the gauge theory lives. This boundary
reflects back the thermal radiation, and the black branes are hence
in stable thermal equilibrium, as opposed to black holes in flat spacetime,
which evaporate away%
\footnote{Also, black holes in our universe tend to have very small temperatures
of order of mK, as black holes of higher temperature are hard to form
and would quickly evaporate. The thermal stability of black holes
in $AdS$ allows any temperature, and in fact the temperature of black
holes representing a quark-gluon plasma would have the same very high
temperature of order $10^{12}$K.%
}.

It is also possible to construct gradient expansion solutions to Einstein's equations,
where the local geometry is approximately that of boosted black brane with slowly varying temperature
and boost velocity \cite{Bhattacharyya:2008jc,Hubeny:2011hd}.  
In this case the corresponding 
dual gauge theory state is that of a system in local thermal equilibrium with dynamics 
governed by hydrodynamics.  Simply put, in the limit of slowly varying fields Einstein's equations 
reduce to hydrodynamics in one less dimension.  The connection between 
slowly varying solutions to Einstein's equations and hydrodynamics is known as the fluid/gravity correspondence.
By matching the gradient expansion obtained by solving Einstein's equations to the constitutive relations 
of hydrodynamics, transport coefficients such as the viscosity can be computed at strong coupling.

At first order in gradients the constitutive relation for the stress tensor $T^{\mu \nu}$ in relativistic neutral hydrodynamics 
read \cite{landau1959fluid}
\begin{eqnarray}
T_{\mu\nu} & = & e\, u_{\mu}u_{\nu}+p[e]\Delta_{\mu\nu}+\pi_{\mu\nu},\text{ where,}\label{eq:hydro-constituive}\\
\Delta_{\mu\nu} & = & g_{\mu\nu}+u_{\mu}u_{\nu}\text{ and }\\
\pi_{\mu\nu} & = & -\eta[e]\,\sigma_{\mu\nu}-\zeta[e]\,\Delta_{\mu\nu}(\nabla\cdot u)+\mathcal{O}(\partial^{2}),\text{ with}\\
\sigma_{\mu\nu} & = & \Delta_{\mu\alpha}\Delta_{\nu\beta}(\nabla^{\mu}u^{\nu}+\nabla^{\nu}u^{\mu})-\frac{2}{3}\Delta_{\mu\nu}\Delta_{\alpha\beta}\nabla^{\alpha}u^{\beta},\label{eq:sigma}
\end{eqnarray}
where $e$ is the proper energy density, $u_{\mu}$ the  fluid
velocity, $p[e]$ is the pressure (which is a function of the proper energy via an equation of state),
 $\pi_{\mu\nu}$ is the shear tensor, $\eta$ the shear
viscosity and $\zeta$ is the bulk viscosity, which vanishes in scale
invariant theories such as SYM. Note that the fluid velocity and energy density
are defined as the time-like eigenvector and associated eigenvalue
of the stress tensor ($T_{\mu\nu}u^{\nu}=-e\, u_{\mu})$ and that $\sigma^{\mu \nu}$
is transverse and traceless: $u^{\mu}\sigma_{\mu\nu}=\sigma_{\,\mu}^{\mu}=0$.
Alternatively one can say that the fluid velocity is defined such
that when boosting $T_{\mu\nu}$ with velocity $u_{\mu}$ there is
no momentum flow, i.e. $T'_{0i}=0$, which is called the Landau frame%
\footnote{When including a charge density with corresponding current and chemical
potential it is also possible to define a frame where there is no
charge flow, which is called the Eckart frame \cite{Eckart:1940te}.
In this review we do not include a charge density.%
}. 
The conservation equation $\nabla^{\mu}T_{\mu\nu}=0$ together with
the equation of state and the viscosity (computed theoretically or
measured) now form a closed system of equations, fully determined
by a given initial energy density and fluid velocity.

The equation of state, viscosity and higher order transport coefficients
are not determined by hydrodynamics itself, and are a property of
the microscopic theory under consideration.  In weakly coupled QCD it is 
possible to use the Boltzmann equation to perturbatively compute the
shear viscosity, which led to \cite{Arnold:2000dr}
\begin{equation}
\eta_{\text{weak}}=\kappa\frac{T^{3}}{g^{4}\log(1/g)}+\mathcal{O}(\frac{1}{g^{4}}),\label{eq:viscosityweak}
\end{equation}
with $T$ the temperature, $g$ the gauge coupling and $\kappa$ a
coefficient depending on the number of colors and flavors, which can
be computed numerically. Non-perturbatively it is unknown how to compute
the viscosity of QCD. However, there have been attempts to compute the viscosity using lattice
techniques (see \cite{Meyer:2011gj} for a review).  It was therefore ground-breaking when a few years after the discovery
of  gauge/gravity duality Policastro, Son and Starinets computed
the shear viscosity of a thermal plasma described by strongly coupled
SYM theory \cite{Policastro:2001yc}: 
\begin{equation}
\eta =\frac{s}{4\pi}=\frac{\pi}{8}N_{c}^{2}T^{3},
\end{equation}
with $s$ is the entropy density. The factor of $1/4\pi$ is surprisingly universal and valid
for any strong coupling and large $N_{c}$ quantum field theory with
a gravitational dual. For instance adding charge density changes both
the shear viscosity and the entropy density, but its ratio is still
$1/4\pi$. 

Nevertheless,
it is possible to have holographic models with different viscosity
to entropy ratios, most notably by including higher derivative corrections
in the Einstein equations. In this case corrections for finite $N_{c}$
can actually make the $\eta/s$ ratio drop below $1/4\pi$, thereby
violating the conjectured universal lower bound \cite{Kovtun:2004de}
on the viscosity \cite{Kats:2007mq}. Nevertheless, general considerations
using the quantum mechanical uncertainty principle still hint at the
existence of a lower bound, even though its value is likely smaller than $1/4\pi$.

Holography can also yield finite coupling and finite $N_c$ corrections to the viscosity.
For SYM the results read  \cite{Buchel:2004di,Buchel:2008sh,Myers:2008yi}:
\begin{equation}
\eta/s=\frac{1}{4\pi}\left(1+\frac{15}{\lambda^{3/2}}\zeta(3)+\frac{5}{16}\frac{\lambda^{1/2}}{N_{c}^{2}}\right),
\end{equation}
with $\lambda=g^{2}N_{c}$ the 't Hooft coupling. As expected from (\ref{eq:viscosityweak})
the viscosity increases as the coupling constant decreases, and the
increase can be significant at reasonable values of the coupling
constant (for $\lambda\sim20$ and $N_{c}\sim3$ the corrections both
are roughly 20\%). Current hydrodynamic models of the quark-gluon
plasma indeed favor an $\eta/s$ in the range $1/4\pi$ to $2.5/4\pi$
\cite{Romatschke:2007mq,Heinz:2011kt,Schenke:2012hg,Gale:2012rq}.

\subsection{\noindent Second order relativistic hydrodynamics}

\begin{figure}
\begin{centering}
\includegraphics[width=6cm]{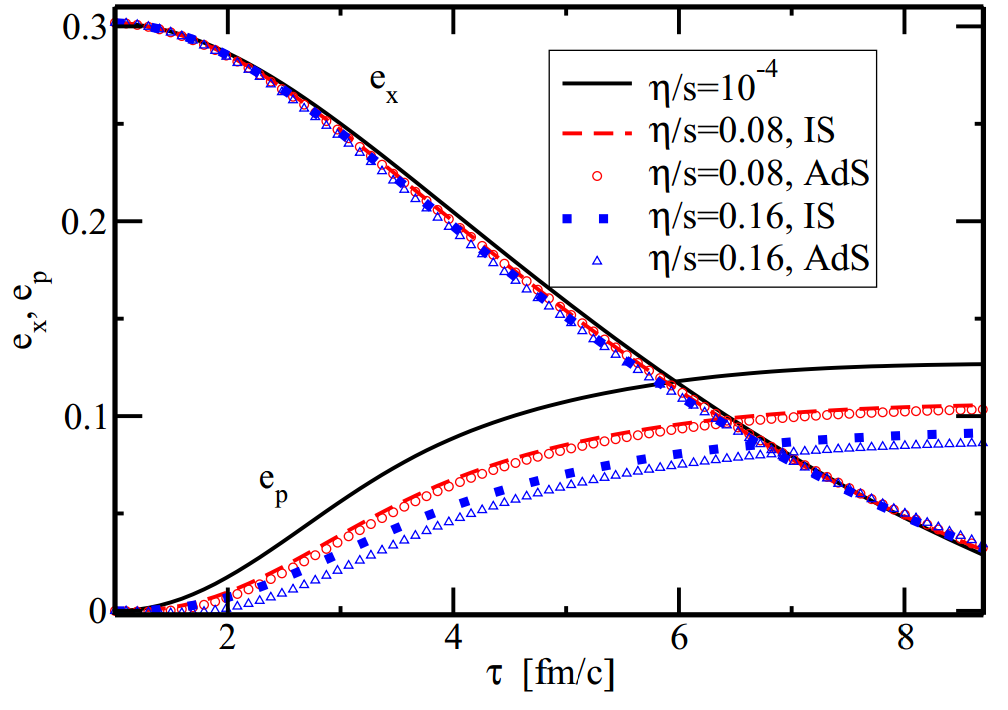}\includegraphics[width=6cm]{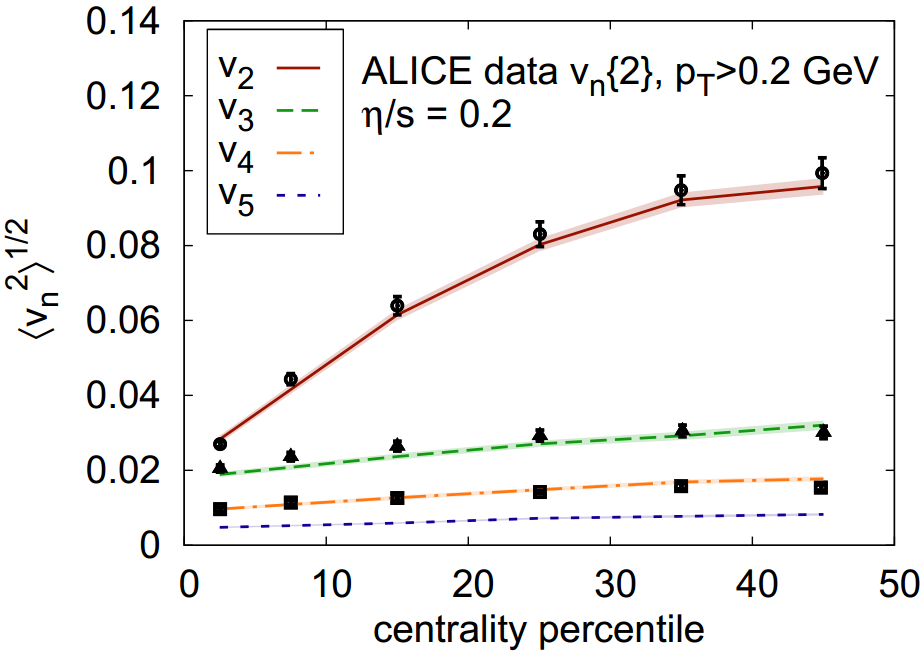}
\par\end{centering}

\protect\caption{(left) The second order hydrodynamics derived using AdS/CFT (AdS)
is compared with the older weakly coupled Muller-Israel-Stewart hydrodynamcis
(IS), which uses $\tau_{\pi}=6\eta/sT$ and does not include the $\lambda_{1}$
or $\lambda_{2}$ in eqn. \ref{eq:2ndorder}. For these initial conditions
the spatial and momentum anisotropy ($e_x$ and $e_p$) do not depend much on the precise value of the second order
transport coefficients (figure from \cite{Luzum:2008cw}). (right)
Similar hydrodynamics has been used to obtain recent viscosity over entropy ratios 
of roughly 0.12 and 0.2 at top RHIC and LHC collisions respectively 
(LHC estimate shown, figure from \cite{Gale:2012rq}).\label{fig:hydros}}
\end{figure}

\noindent Holography and in particular the fluid/gravity duality also
has led to a more systematic understanding of (relativistic) hydrodynamics
itself \cite{Baier:2007ix,Bhattacharyya:2008jc,Loganayagam:2008is}.
As alluded to above, via the fluid/gravity correspondence it is possible to systematically 
solve Einstein's equations.  The gradient expansion solution
encodes the constitutive relations of hydrodynamics to \textit{all orders} in gradients
and associated transport coefficients \cite{Baier:2007ix,Bhattacharyya:2008jc}.
Let us begin by simply stating the results of \cite{Baier:2007ix,Bhattacharyya:2008jc}
for the shear tensor $\pi^{\mu \nu}$ up to the second order in gradients.

Since SYM is conformal, to write $\pi^{\mu \nu}$ at second order in gradients it is useful to employ derivative operators $\mathcal{D}_\mu$ which 
transform covariantly under conformal transformations.  For tensors $Q_{\nu\ldots}^{\mu\ldots}$ which, under a conformal transformation $g_{\mu\nu}\rightarrow e^{2\phi}\tilde{g}_{\mu\nu}$,
transform like $Q_{\nu\ldots}^{\mu\ldots}=e^{-w\phi}\widetilde{Q}_{\nu\ldots}^{\mu\ldots}$, $\mathcal D$ is defined by
\cite{Loganayagam:2008is}
\begin{align*}
\mathcal{D}_{\sigma}Q_{\nu\ldots}^{\mu\ldots} & \equiv\partial_{\sigma}Q_{\nu\ldots}^{\mu\ldots}+w\,\mathcal{A_{\sigma}}Q_{\nu\ldots}^{\mu\ldots}+\gamma_{\sigma\lambda}^{\mu}Q_{\nu\ldots}^{\lambda\ldots}+\ldots-\gamma_{\sigma\nu}^{\lambda}Q_{\lambda\ldots}^{\mu\ldots}-\ldots,\\
\gamma_{\mu\nu}^{\lambda} & \equiv g_{\mu\nu}A^{\lambda}-\delta_{\mu}^{\lambda}A_{\nu}-\delta_{\nu}^{\lambda}A_{\mu},\\
A_{\mu} & \equiv u^{\nu}\partial_{\nu}u_{\mu}-\frac{\partial_{\nu}u^{\nu}}{3}u_{\mu}.
\end{align*}
With this definition derivatives transform covariantly: $\mathcal{D}_{\lambda}Q_{\nu\ldots}^{\mu\ldots}=e^{-w\phi}\mathcal{D}_{\lambda}\widetilde{Q}_{\nu\ldots}^{\mu\ldots}$, and for instance Eq.~(\ref{eq:sigma}) simplifies  to $\sigma_{\mu\nu}=\mathcal{D}_{\mu}u_{\nu}+\mathcal{D}_{\nu}u_{\mu}$.
The full second order conformal shear tensor is now given by%
\footnote{Most of these computations have been done for fluids living in a curved
spacetime \cite{Baier:2007ix}, which gives an extra term not considered
in this review.%
}:
\begin{eqnarray}
\pi_{\mu\nu} & = & -\eta\,\sigma_{\mu\nu}+\eta\,\tau_{\pi}\, u^{\lambda}\mathcal{D}_{\lambda}\sigma_{\mu\nu}+\lambda_{1}[\sigma_{\mu\lambda}\sigma_{\,\nu}^{\lambda}-\frac{\Delta_{\mu\nu}}{3}\sigma^{\alpha\beta}\sigma_{\alpha\beta}]\nonumber \\
 &  & +\lambda_{2}[\omega_{\mu\lambda}\sigma_{\,\nu}^{\lambda}+\omega_{\nu\lambda}\sigma_{\,\mu}^{\lambda}]+\lambda_{3}[\omega_{\mu\lambda}\omega_{\,\nu}^{\lambda}+\frac{\Delta_{\mu\nu}}{3}\omega^{\alpha\beta}\omega_{\alpha\beta}],\label{eq:2ndorder}
\end{eqnarray}
where $\omega_{\mu\nu}=\mathcal{D}_{\mu}u_{\nu}-\mathcal{D}_{\nu}u_{\mu}$.
In strongly coupled SYM the second order transport coefficients
are given by $\tau_{\pi}=(2-\ln(2))/2\pi T$, $\lambda_{1}=\eta/2\pi T=2\lambda_{2}/\ln(2)$
and $\lambda_{3}=0$ \cite{Bhattacharyya:2008jc}. Note that all these
coefficients get considerably more complicated when including a chemical
potential; in particular $\lambda_{3}$ becomes non-zero\cite{Erdmenger:2008rm}.

Second order hydrodynamics is particularly useful for numerical simulations
of viscous hydrodynamics. This is because high momentum
modes in first order relativistic hydrodynamics can propagate faster than the speed of light.
Fundamentally this is not a problem, as hydrodynamics
is an expansion in long wavelengths, i.e. small momenta, and these
acausal modes are outside the regime of its applicability \cite{Geroch:1995bx}.
Nevertheless, these modes make the equations numerically unstable
\cite{Hiscock:1985zz} and therefore in numerical simulations a second
order term was added to stabilize the equations, giving Muller-Israel-Stewart
hydrodynamics \cite{Muller:1967zza,Israel:1979wp}.

Muller-Israel-Stewart hydrodynamics can be seen as a special case
of second order hydrodynamics where $\lambda_{i}=0$. 
One insight which holography helped solidify is that, in contrast to the Muller-Israel-Stewart theory, at a given order in the gradient expansion
one must add every possible tensor structure to the constitutive relations which is consistent with the  symmetries  of the underlying quantum field theory.
This had lead to the 
development of causal viscous relativistic hydrodynamics \cite{Baier:2006gy,Luzum:2008cw,Romatschke:2009im}. 
In conformal theories such as SYM the equations of second order hydrodynamics give causal propagation for high momenta modes.

While the second order transport coefficients make simulations much easier to
perform, the precise value of these coefficients fortunately does
not seem to have a big influence on the particle spectra in realistic
simulations (see figure \ref{fig:hydros}, left). The latter feature
does however crucially depend on using conformal hydrodynamic equations;
when using (second order) non-conformal equations results do seem
to depend on the second order coefficients \cite{Song:2007fn}. One
of the biggest successes of these equations are estimates for the
shear viscosity of the quark-gluon plasma \cite{Romatschke:2007mq,Heinz:2011kt,Schenke:2012hg,Gale:2012rq},
as shown in Fig.~\ref{fig:hydros} (right).

Another more recent example of a better understanding of hydrodynamics
comes from \cite{Heller:2013fn}, where in a specific boost-invariant
setting all transport coefficients were computed numerically up to
order 240 in the gradient expansion. With those coefficients it could
be shown that the hydrodynamic gradient expansion is not necessarily
convergent, it was found to be an asymptotic series. Moreover, it
was understood why hydrodynamics did not converge, since the divergent
hydrodynamic expansion contains information about the non-hydrodynamic
modes. Within AdS/CFT these modes are included in the so-called quasi-normal
modes: vibrations of the black hole horizon. Strikingly, by a Borel
resummation reference \cite{Heller:2013fn} was able to extract the
precise value of the dominant quasi-normal mode from the hydrodynamic
expansion, thereby showing how hydrodynamics contains information
about its own break-down.

Much work has been devoted to study charged hydrodynamics, extending
the previous formulas to a plasma with a conserved current, chemical
potential and various new transport coefficients, see for instance
\cite{Erdmenger:2008rm,Banerjee:2008th}. More recently there has
been interest in anomalous hydrodynamics, where quantum anomalies
lead to modified hydrodynamic equations. A plasma with a non-trivial
axial charge density will develop an electric current $\vec{J}$ in
the direction of an external magnetic field $\vec{B}$ (chiral magnetic
effect or CME) \cite{Kharzeev:2007tn,Kharzeev:2013ffa}:
\begin{equation}
\vec{J}=\frac{e^{2}}{2\pi^{2}}\mu_{5}\vec{B},
\end{equation}
where $\mu_{5}$ is the chemical potential associated with the axial
charge. For averaged heavy ion events $\mu_{5}$ would be zero, but
a full event-by-event anomalous hydrodynamic simulation suggests that
this effect is measurable in heavy ion collisions \cite{Hirono:2014oda}.
Similarly to the analysis above, also for the CME holography has proved
extremely useful to show how this effect is in fact necessarily included
already in first order charged hydrodynamics, and also to estimate the relevant transport coefficients
at strong coupling \cite{Son:2009tf,Gynther:2010ed}.

Lastly, inspiration from the symmetries present in AdS has led to non-trivial analytic solutions of the hydrodynamic equations.  An interesting solution was provided by Gubser \cite{Gubser:2010ze}, where a solution to viscous conformal hydrodynamics with boost-invariance and a non-trivial expansion in the transverse plane was found.
This solution follows from a simple set-up in AdS, but then leads to this non-trivial analytic and hence very useful hydrodynamic solution. Recently, this approach has been extended to an exact  solution of the relativistic Boltzmann equation \cite{Denicol:2014xca}.

Interestingly, aside from the computation of transport coefficients, much of the development in hydrodynamics above need
not rely on gauge/gravity duality. Nevertheless, inspiration
from the duality has led to this more precise understanding of hydrodynamics,
by using gauge/gravity as a theoretical arena to get new insights.

\section{Fast thermalization and the resulting stress tensor\label{sec:Fast-thermalization-and}}

\subsection{Simple models}

\noindent Perhaps an even more remarkable feature of holography is
the straightforward possibility to study far-from-equilibrium dynamics in strongly coupled gauge theories.
How do non-equilibrium systems  thermalize and relax to local equilibrium?
Using holography answering this question amounts
to studying real time dynamics in the gravitational dual, which is
not much different from conventional problems in general relativity,
such as binary neutron stars, or black hole inspirals \cite{Lehner:2001wq,Pretorius:2005gq}.
As is well-known these latter problems can still be very challenging
from a numerical point of view, but compared to non-perturbative quantum
field theory it is definitely an enormous simplification.  For a review of the use of numerical relativity 
in holography see for example \cite{Chesler:2013lia}.

The first studies of holographic thermalization computed how perturbed black holes
relax to homogeneous black holes, computing the so-called quasi-normal
modes \cite{Horowitz:1999jd}. These are determined by expanding the
perturbation of the metric in a sum of modes proportional to $e^{-i\omega_{i}t}$,
with complex frequency $\omega_{i}$. The Einstein equations and boundary
conditions at the horizon and boundary then dictate that there is
a discrete spectrum, whereby the mode always decays (i.e. $\Im(\omega_{i})<0$).
The mode with the slowest decay will then determine the late-time
decay back to the thermal black hole, and hence can be said to determine
the thermalization time in this setting. Dimensional analysis in this
scale invariant theory dictates that $\omega_{i}\sim T$ , but the
coefficient has to be computed numerically, giving $\Im(\omega_{0})\approx8.6 T$
\cite{Horowitz:1999jd}. 

The estimate from quasi-normal modes could imply a thermalization
time $t_{\rm therm} <1/T$, which for a temperature of 1000 MeV would be about
0.2 fm/c. Compared to perturbative methods this is very fast, which
should be expected for a system at strong coupling. On the other hand,
even though the coupling is assumed to be infinite, there is still
a finite answer for the thermalization time, leading to the intuition
that strong interactions still have some microscopic physics present as the scale $1/T$.

\begin{figure}
\begin{centering}
\includegraphics[width=6cm]{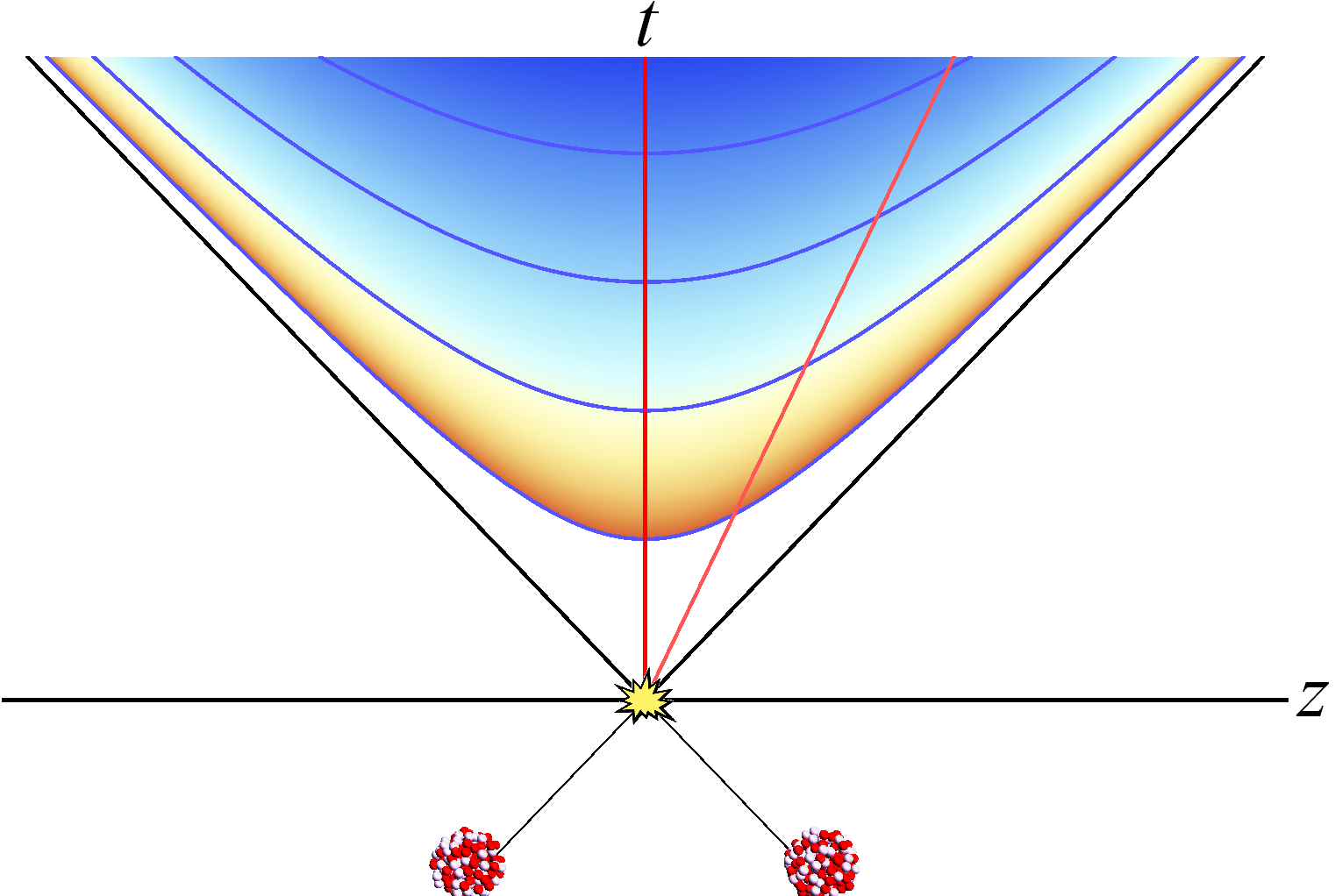}
\par\end{centering}

\protect\caption{A simple and often used model of a heavy-ion collision was proposed
by Bjorken \cite{Bjorken:1982qr}, where he assumed that a heavy-ion
collision is approximately boost-invariant, at least near $z=0$.
This means that all physics only depends on proper time $\tau=\sqrt{t^{2}-z^{2}}$,
and hence that all physics experienced by observers, such as a temperature
field illustrated here, is independent of its frame. The two red lines
would illustrate two such frames, which indeed has equal temperatures
at equal $\tau$ (figure from \cite{vanderSchee:2014qwa}).\label{fig:boost-invariance}}
\end{figure}

Nevertheless, the the quasi-normal mode analysis is only valid near equilibrium:
the state is assumed to be a small perturbation of the final thermal
state. In heavy ion collisions the initial state is very far-from-equilibrium, 
so solving the full non-linear Einstein equations is
of great importance. One of the first studies solving this numerically
is presented in \cite{Chesler:2008hg,Chesler:2009cy}, where quark-gluon plasma formation 
and thermalization was studied in both homogeneous and boost invariant set-ups (see fig.
\ref{fig:boost-invariance}).   The gauge theory dynamics was, by symmetry, effectively $1+0$ dimensional
and the dual gravitational dynamics was $1+1$ dimensional.
In these papers non-equilibrium states were created 
by starting in the ground state and turning on background fields.  
It was found that even in the case where the initial state is the vacuum, a thermal state can 
be achieved in times of order $1/T_{}$ with $T_{}$ the final state temperature.

An intriguing finding in \cite{Chesler:2009cy} was that when the system first starts 
to behave hydrodynamically, viscous effects were order one: the first order
viscous correction to the constitutive relations (\ref{eq:hydro-constituive}) is as big as the leading term. 
Nevertheless, even with such large viscous corrections the hydrodynamic gradient expansion was seen
to be well behaved.  Moreover, the applicability of hydrodynamics was seen not to be governed by convergence 
of the hydrodynamic gradient expansion.  Instead, the applicability of hydrodynamics was seen to be governed by
the decay of non-hydrodynamic degrees of freedom.

The approach in \cite{Chesler:2009cy} was extended in \cite{Heller:2011ju}.
Here the main feature was the possibility to chose several initial states and to evolve them without turning on any background fields.
For all initial states they found a thermalization time less than $1/T$, with $T$ the temperature at the moment of thermalization. 

\begin{figure}
\begin{centering}
\includegraphics[width=5.80cm]{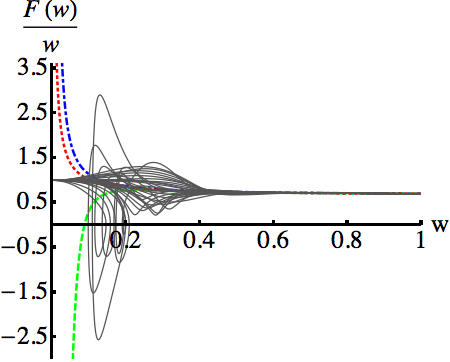}\includegraphics[width=7.2cm]{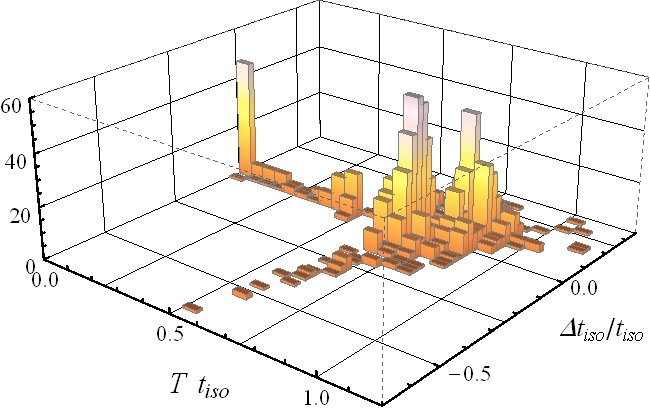}
\par\end{centering}

\protect\caption{The left figure shows boost invariant thermalization, where the evolution
of 29 random initial states of $F(w)\equiv\frac{\tau}{w}\partial_{\tau}w$
as a function of $w\equiv T_{eff}\tau$ is compared with predictions
from first-, second- and third-order hydrodynamics (figure from \cite{Heller:2011ju}).
It can be seen that they all thermalize in a time where $\tau\lesssim0.5/T_{eff}$,
where $T_{eff}$ is the temperature associated with the energy density
at that time. The right figure did a similar analysis in a homogeneous
set-up, using over 1000 initial states. Also there thermalization
always happens within $t\lesssim1/T$. Here it is also shown that
the difference between the linearized and fully non-linear Einstein
equations is small, almost always less than 20\% (figure from \cite{Heller:2012km}).
\label{fig:simple-thermalisation}}
\end{figure}

Of particular utility, this paper also introduced a dimensionless quantitiy,  $F(w)\equiv\frac{\tau}{w}\partial_{\tau}w$, where $w\equiv T_{eff}\tau$ can be thought of as dimensionless time and $T_{eff}$  is the temperature associated with the energy density
at that time. The dimensionless function $F(w)$ is completely determined within hydrodynamics and for instance equals 2/3 for ideal hydrodynamics. As $F(w)$ is unique within hydrodynamics it is clear that differences between $F(w)$ found for the different evolutions in AdS/CFT must be due to genuine non-hydrodynamic effects, as is convincingly shown on the left of  Fig. \ref{fig:simple-thermalisation}.

The right of figure \ref{fig:simple-thermalisation} shows a similar
study in a homogeneous setting, with more than 1000 initial states,
also leading to thermalization within a time of $1/T$ \cite{Heller:2012km,Heller:2013oxa}.
The main new result in this paper was to compare the full non-linear
thermalization process with the linearized quasi-normal mode approach
outlined in the second paragraph of this section. Quite surprisingly
it was found that for general initial states the evolution was always
well described, within 20\%, by the linearized approximation. This
then also confirmed that the quasi-normal mode estimates of strongly
coupled thermalization times were indeed accurate also at the non-linear
level.

Lastly, these studies can be corrected for finite coupling effects,
which has recently been done in \cite{Steineder:2012si,Stricker:2013lma}.
Not surprisingly, finite coupling corrections slow down the thermalization,
as in these papers becomes apparent through a smaller imaginary part
of the lowest quasi-normal mode.

\subsection{\noindent Colliding shock waves\label{sub:Colliding-shock-waves}}

\noindent All models presented above contained enough spacetime symmetry to make the gauge theory 
dynamics effectively $1+0$ dimensional.  Here we shall review results where the gauge theory dynamics are allowed to be 
$1+1$ and $1+2$ dimensional.  In holography
the energy-momentum in the gauge theory directly corresponds to the
metric in the gravitational theory, so to simulate highly energetic
colliding nuclei it is therefore natural to consider collisions of
gravitational shock waves, moving at the speed of light \cite{Janik:2005zt,Albacete:2008vs,Gubser:2008pc,Grumiller:2008va,Lin:2009pn,Aref'eva:2009wz}.
The shock waves can be viewed as analogous to the shock waves constructed in
flat space by Dray and 't Hooft \cite{Dray:1984ha}, which consist
of taking a Schwarzschild black hole and boosting this to the speed
of light, keeping its total energy fixed.  The shock waves are dual to lumps of energy in the gauge theory, which propagate at the speed of light.
When the shock waves collide they can produce a black hole which will then
relax to equilibrium.  Likewise, in the dual gauge theory, the collision will result in the formation of a non-Abelian plasma which 
will subsequently thermalize.

One interesting result from the aforementioned studies of shock wave collisions was the entropy production,
estimated by the area of the trapped surface of the formed black hole
at the time of collision. This trapped surface gives a lower bound
on the area of the black hole horizon, which is directly related to
the produced entropy in the gauge theory. Ref. \cite{Gubser:2008pc} estimated
that:
\begin{equation}
S\geq S_{trapped}\approx35000\left(\frac{\sqrt{s_{NN}}}{200\,\text{GeV}}\right)^{2/3},\label{eq:entropy}
\end{equation}
where $S$ is the entropy produced in a central gold nucleus-nucleus collision with a nucleon-nucleon
collision of energy $\sqrt{s_{NN}}$ and $S_{trapped}$ is the entropy
of the trapped surface. The entropy increase during the evolution
of the quark-gluon plasma is not too big due too to the relatively
small viscosity, but can in fact be sizable as highly anisotropic
plasmas increase the entropy production. Nevertheless Eq.~(\ref{eq:entropy})
gives a lower bound on the final entropy produced, which is directly
related to the measurable total charged particle multiplicity $N_{tot,ch}=S/7.5$ \cite{Muller:2005en}. In \cite{vanderSchee:2014qwa}
the estimate (\ref{eq:entropy}) was made somewhat stronger by doing
calculations numerically till later times, but two important features
remained. Firstly the numerical value is about right for RHIC collisions,
albeit on the high side, which is impressive as Eq.~(\ref{eq:entropy})
is a direct calculation, without having any input from other experiments.
Secondly, the $2/3$ power law as a function of $\sqrt{s_{NN}}$ is
a robust outcome of a scale invariant theory. Nevertheless, with the
LHC results now known \cite{Collaboration:2011rta}, the experimental
data favors a power of $1/2$, and Eq.~(\ref{eq:entropy}) therefore
overestimated the multiplicity at LHC collisions. This can be taken
as an indication that QCD is not scale invariant at energies probed
between RHIC and LHC.

The previous calculations could not say much about the dynamics after
the collision, which requires solving the full non-linear Einstein
equations. It was therefore a major breakthrough when a real collision
could be simulated within gravity \cite{Chesler:2010bi}. This study
collides planar gravitational shock waves, thereby simplifying the
problem by neglecting transverse dynamics. The initial conditions
for the shock wave are determined by the stress tensor
\begin{equation}
T_{\pm\pm}(z_{\pm})=\kappa\, h(z_{\pm}),\label{eq:shocks}
\end{equation}
where $z_{\pm}=t\pm z$ are null coordinates along the collision axis 
and $\kappa=N_{c}^{2}/2\pi^{2}$ is
a measure of the degrees of freedom of the gauge theory. All other
components of the stress tensor are zero, and within pure gravity
this fully determines the metric before the collision. For this review
we restrict $h$ to be Gaussian: 
\begin{equation}
h(z)=\frac{\mu^{3}}{w\sqrt{2\pi}}\exp\left[-\frac{z^{2}}{2w^{2}}\right],\label{eq:choiceforh}
\end{equation}
where $w$ is the width of the Gaussian and $\kappa\,\mu^{3}=\kappa\int_{-\infty}^{\infty}h(z)dz$
is the total energy per transverse area. 

In what follows we make the following choices for parameters.
We choose $N_{c}=1.8$ such that the ratio $e/T^{4}\approx12$ matches
lattice QCD simulations \cite{Gubser:2008pc}.  The choice of $N_c < 3$ compensates 
for the fact that SYM has approximately three times as many degrees of freedom 
than QCD.
We then fix $\mu$ by setting the 
\begin{equation}
\kappa\,\mu^{3}\approx\frac{\sqrt{s_{NN}}}{2}\frac{N_{\text{Pb}}}{\pi R^{2}},
\end{equation}
where $N_{\text{Pb}}=206$ and $R=6$ fm.  The right hand side of this equation 
is simply the energy density per unit area near the center of a lead nucleus.
We choose the width of the Gaussian to be
$w\approx\frac{4}{3}\, R\, m_{N}/\sqrt{s_{NN}}$,
with $m_N$ the nucleon mass.
Of course real nuclei are not Gaussians, (nor is QCD $=$ SYM) 
so all these numbers have to be taken with a grain
of salt. Nevertheless, these choices gives a qualitative understanding of the scales in the problem.

\begin{figure}
\centering{}
\includegraphics[width=6.6cm]{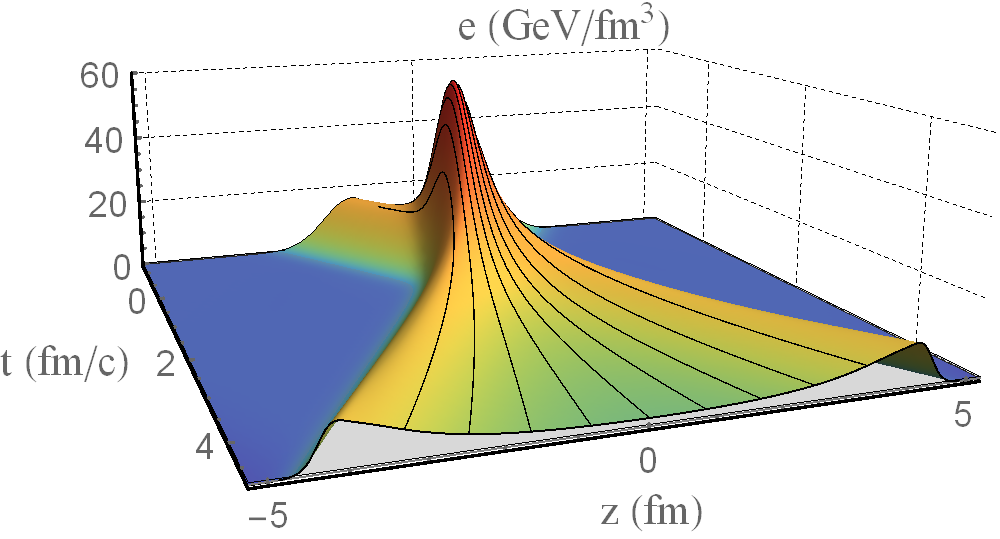}\includegraphics[width=6.6cm]{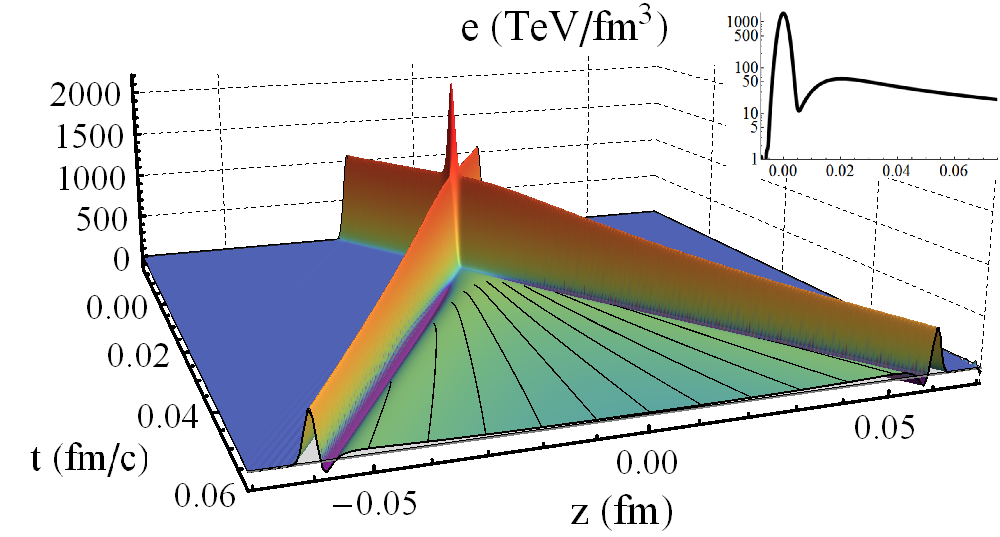}
\protect\caption{Energy density $e$ for collisions of shock waves at low energy ($\sqrt{s_{NN}}=19.3\,\text{GeV}$,
left) and high energy ($\sqrt{s_{NN}}=2.76\,\text{TeV}$, right) as
a function of time and longitudinal coordinate $z$. The \foreignlanguage{british}{grey} planes
lie at the origin of the vertical axes. Note that these simulations
are planar in the transverse plane and therefore neglect transverse
dynamics. For the short time scales presented at the right plot this
is probably a good approximation, but the simulation on the left will
change when taking transverse expansion into account as well (figures
adapted from \cite{Casalderrey-Solana:2013aba}).\label{fig:EnergyDensity} }
\end{figure}

After setting up these initial conditions it is possible to numerically
solve the Einstein equations (see \cite{Chesler:2013lia,vanderSchee:2014qwa}
for an introduction), and thereafter extract the expectation value
of the gauge theory stress tensor. Figure \ref{fig:EnergyDensity}
shows the energy density obtained
in this way for collisions having $\sqrt{s_{NN}}=19.3\,\text{GeV}$
and $\sqrt{s_{NN}}=2.76\,\text{TeV}$. For a proper comparison with
\cite{Chesler:2010bi,Casalderrey-Solana:2013aba} one should notice
that the dimensionless product 
\begin{equation}
\mu w=\left(\frac{\sqrt{s_{NN}}}{2\kappa}\frac{N_{lead}}{\pi R^{2}}\right)^{1/3}\frac{4}{3}\, R\, m_{N}/\sqrt{s_{NN}}\approx23(\sqrt{s_{NN}}/\text{GeV})^{-2/3},
\end{equation}
so that $19.3\,\text{GeV}$ and $2.76\,\text{TeV}$ correspond to
a $\mu w$ of 3.2 and 0.11 respectively, whereby we stress once more
that these are ballpark figures, which we think are helpful for a
useful comparison of gauge/gravity results to actual heavy ion
collisions. 

\begin{figure}
\begin{centering}
\includegraphics[width=6cm]{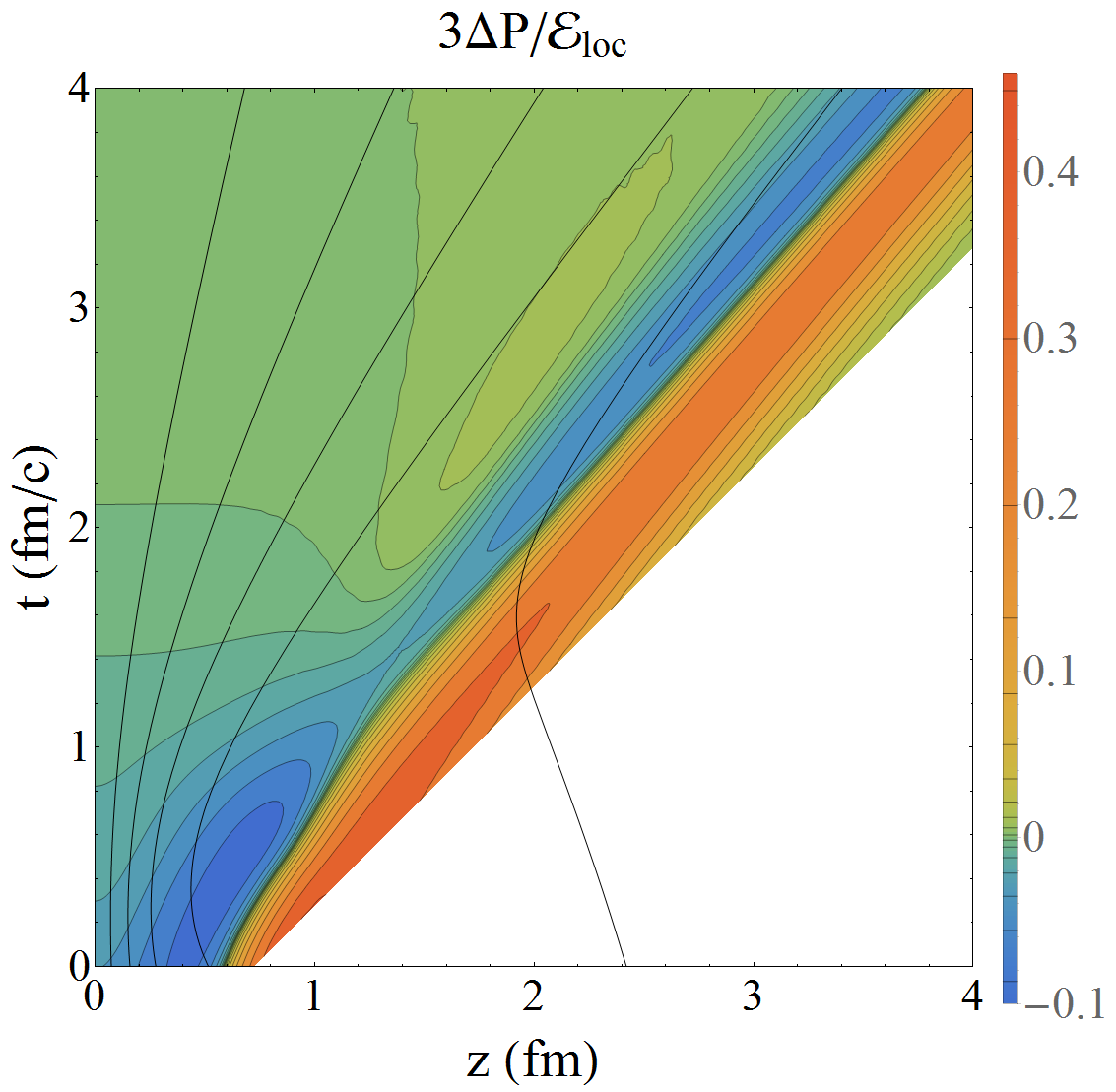}\includegraphics[width=6cm]{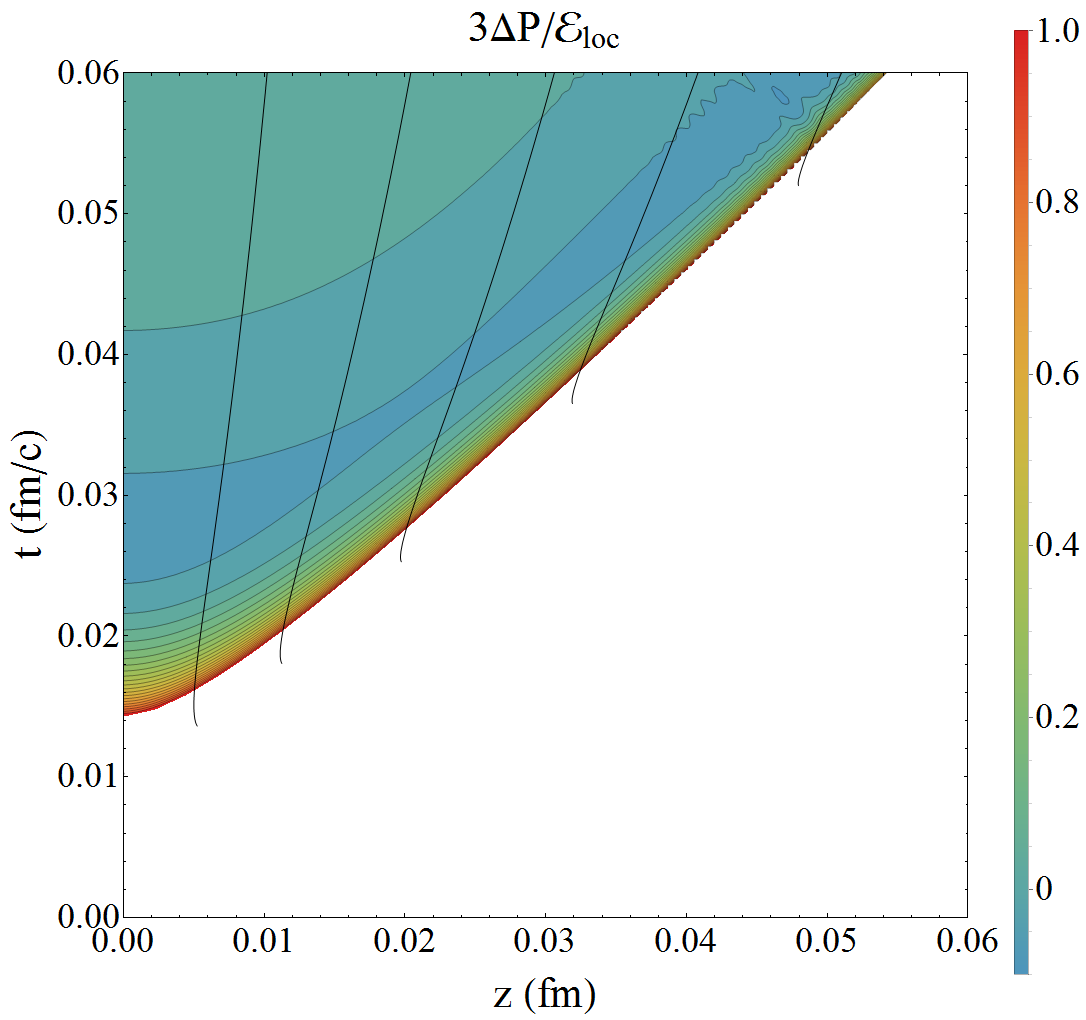}
\par\end{centering}

\centering{}\protect\caption{$3\Delta{\cal P}_{L}^{\textrm{loc}}/\mathcal{E}_{\text{loc}}$ for
low (left) and high (right) energy shocks, with $\Delta P$ the difference
between the longitudinal pressure from holography, and the one computed
using eqn. \ref{eq:hydro-constituive}. The white areas indicate regions
outside the light cone or where hydrodynamics deviates by more than
100\%. The black lines are again stream lines, which for the high
energy shocks stop at the region where no local rest frame exists
\cite{Arnold:2014jva}. \label{fig:hydro-shocks}}
\end{figure}

Several curious features are visible in the resulting energy density
\cite{Casalderrey-Solana:2013aba}. Firstly, for the low energy collisions
the peak energy density is about 60\% higher than the sum of the incoming
energies. This is interpreted as a `pile-up' of energy, as the shocks
thermalize during the collision. For high energy shocks there is not
enough time for this effect, and the shocks therefore only lose there
energy gradually to the plasma, as well as forming a transient region
of negative energy density, where it is not even possible to define
a local rest frame \cite{Arnold:2014jva}. 
Curiously, for the
high energy shocks there seems to be a reasonably accurate analytic formulation --- at least at small rapidities --- 
by using a complexified boost symmetry, even valid in the far-from-equilibrium
regime \cite{Gubser:2014qua}. 
At later times the low and high energy
collisions behave similarly though; they both thermalize fast (shown
in figure \ref{fig:hydro-shocks}) and deposit all their energy into
the plasma.

\begin{figure}
\centering{}\includegraphics[width=6cm]{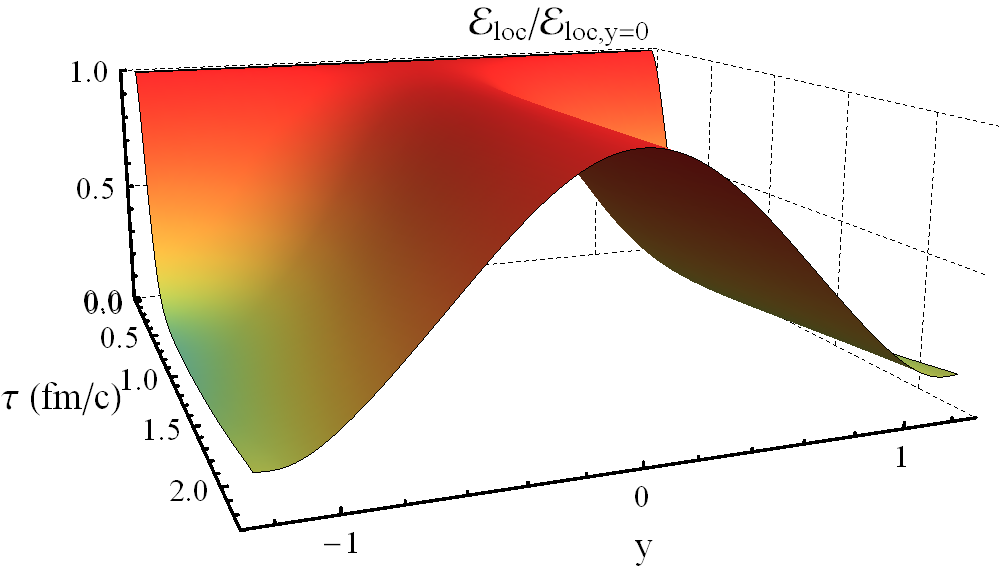}
\includegraphics[width=6cm]{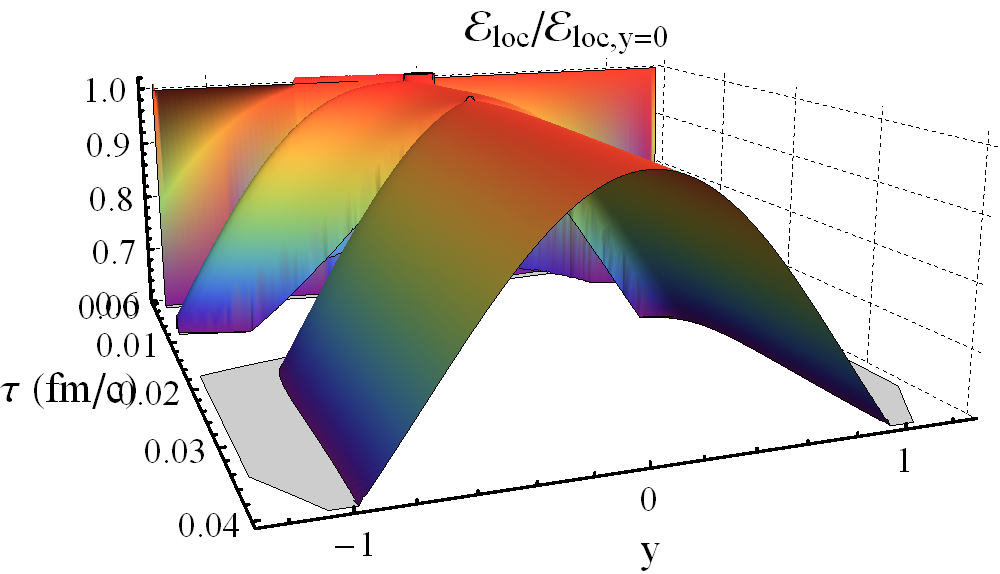}\protect\caption{Energy density in the local rest frame around mid-rapidity as a function
of spacetime rapidity $y$ and proper time $\tau$ for thick (left)
and thin (right) shocks (the same shocks as figure \ref{fig:EnergyDensity}).
In the right case we have excluded from the plot the region in which
the local rest frame is not defined because $2|s|>|e+P_{L}|$ (figures
adapted from \cite{Casalderrey-Solana:2013aba}). \label{fig:rapidity} }
\end{figure}

Interestingly, the formed plasma is quite insensitive to the profile
of the initial shocks, provided characteristic length of the profile
is not too big \cite{Casalderrey-Solana:2013sxa}. This makes the
local energy density of the resulting plasma robust at least for thin
and hence high energy shocks. The rapidity profile of these collisions may even be universal at
strong coupling, and is therefore of considerable interest (figure
\ref{fig:rapidity}, right). The left part of that figure displays
the similar profile for low energy collisions, which looks similar
in shape, but with a width that becomes smaller as the collision energy
decreases.

\subsection{\noindent Including transverse dynamics}

\noindent Lastly, we would like to mention that it is also possible
to include transverse dynamics. Transverse expansion is essential
during the late time of the QGPs explosion, but also early time initial
velocities in the transverse plane influence the transverse momenta
spectra. Nevertheless, a full shock wave collision with with transverse
expansion requires a 3+1 dimensional code in AdS spacetime, as there
is always the extra AdS direction. As this can be technically complicated
it is therefore natural to first try to include transverse dependence
while assuming boost invariance in the longitudinal direction.

Assuming boost invariance is a bit subtle, as Fig.~\ref{fig:rapidity}
convincingly shows a significant dependence on rapidity. Also, using
proper time as a variable usually necessitates to start the gravitational
simulation a finite time after the collision (see \cite{Heller:2012je}
for a counter example though), which has to be accompanied by an assumption
on the dynamics before this time (early time expansion in the case to be presented).
Nevertheless, starting the simulation at such a finite time with several
simple profiles gave two lessons \cite{vanderSchee:2012qj,vanderSchee:2013pia}.
Firstly the plasma again thermalized fast, at a similar time scale
as the examples above. Secondly, after the thermalization the transverse
velocity was given to a good approximation by the following approximation
\cite{vanderSchee:2013pia,Habich:2014jna}:
\begin{equation}
v_{i}=-0.33\tau\,\partial_{i}e/e,
\end{equation}
with $\tau$ the proper time, and $e$ the local energy density in
the transverse plane. This formula is in spirit similar to the universal
pre-flow found in \cite{Vredevoogd:2008id}, but note that only the
flux is universal in the sense of that paper, whereas here we are
interested in the local fluid velocity, which depends on the pressures
as well.

\begin{figure}[t]
\centering{}\includegraphics[width=7.5cm]{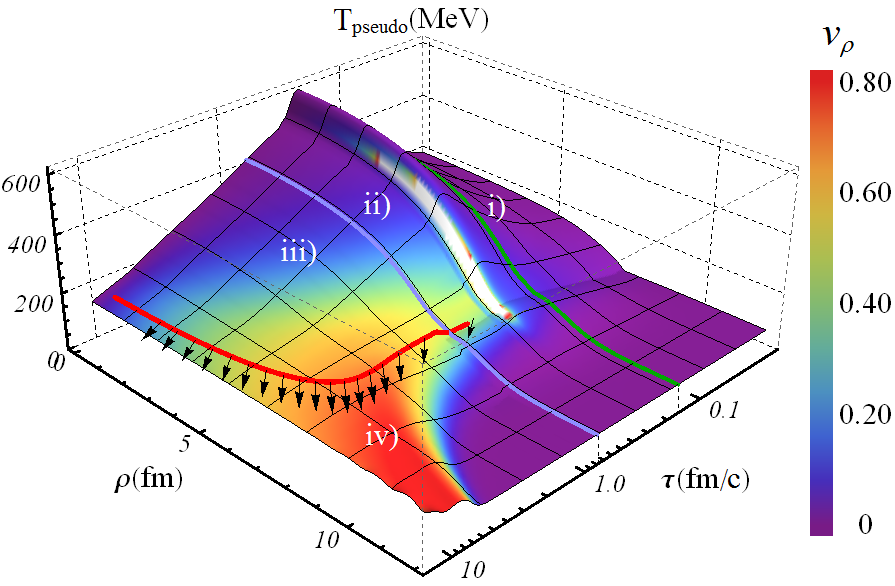}\includegraphics[width=6cm]{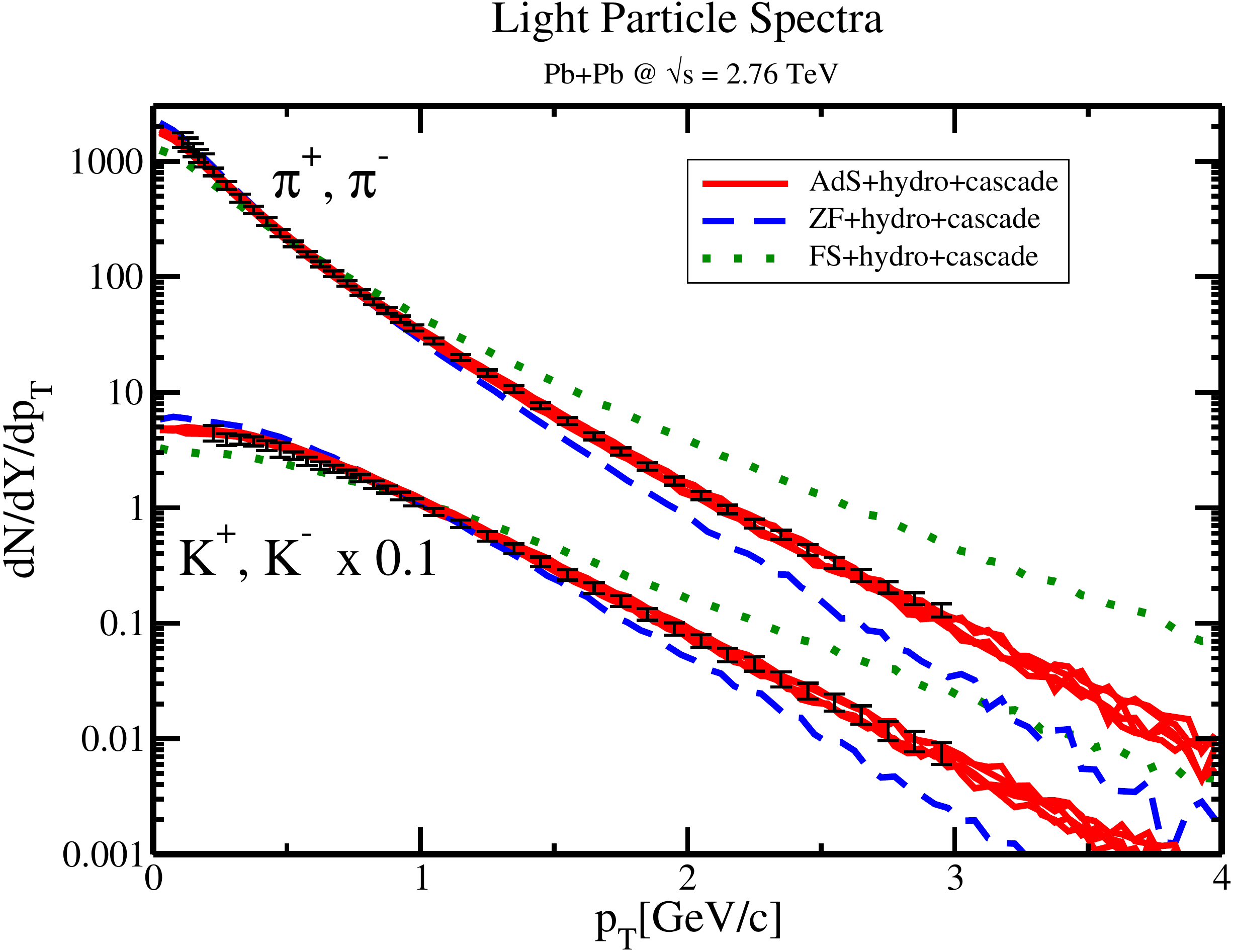}
\protect\caption{From the local energy density we plot the temperature (assuming equilibrium,
therefore labeled $T_{\text{pseudo}}$) and the radial velocity $v_{\rho}=u_{\rho}/u_{\tau}$
for a representative simulation. The plot illustrates four physical
tools used: i) early time expansion, ii) numerical AdS evolution,
iii) viscous hydrodynamics until $T=0.17\,\text{GeV}$, iv) kinetic
theory after conversion into particles (indicated by arrows). The
right plot shows the resulting particle spectra for pions and kaons,
and also includes curves for a zero flow model (ZF) with $v_{i}=0$
until $\tau=1.0$ fm/c, and a free streaming model (FS), with zero
longitudinal pressure (figure from \cite{vanderSchee:2013pia}, experimental
data from ALICE \cite{Abelev:2012wca}). \label{fig:Temp1} }
\end{figure}

Using these simulations it was then possible to couple this to relatively
standard hydrodynamic and cascade codes, and thereafter obtain the
full transverse momenta spectra at mid-rapidity for a central collision,
which fits data surprisingly well (figure \ref{fig:Temp1}). Recently
this has also been extended to several other systems, most notably
also computing the measurable HBT radii \cite{Habich:2014jna}.

\section{Jet energy loss\label{sec:Jet-energy-loss}}

\subsection{Are there jets at strong coupling?}

During the early stages of heavy-ion collisions --- in which the energy density is the highest --- energetic partons 
can be created via hard processes.  The resulting excitations --- jets --- consist of collimated sprays of energy and can traverse the fireball while depositing energy
and momentum into the medium.  Analysis of the particle correlations in the produced jets can provide useful information
about the dynamics of the plasma including the rates of energy loss and momentum broadening \cite{CasalderreySolana:2006sq,Casalderrey-Solana:2014bpa,Casalderrey-Solana:2014wca,Adare:2006nq}.

Why study jets and energy loss at strong coupling?  One piece of evidence that the quark-gluon plasma produced in 
heavy-ion collision is strongly coupled comes from jet quenching itself \cite{Adler:2005ee,Chatrchyan:2011sx,Aad:2010bu}.  Jets traversing the fireball appear to lose their energy very 
quickly, indicating strong interactions with the medium. 

A natural process by which jets can be created is the decay
of off-shell particles such as photons.  At weak coupling the decay 
leads to well-collimated sprays of energy with the polarization of the photon
imprinting itself in the final state in a so called ``antenna pattern" (see for example \cite{Giele:2007di} and references therein).
However, at strong coupling the situation is very different: the decay of off-shell photons 
leads to states which are spherically symmetric (in the rest frame of the off-shell photon): 
all correlation functions at spatial infinity are spherically symmetric and there are no well collimated sprays of energy \cite{Hofman:2008ar}.

What does this mean for studying jets at strong coupling?  Does the observed isotropy mean there are no jets at strong coupling?
Where does the isotropy come from?  
In other words, at what stage in the evolution does the isotropy develop?  Does it come from the underlying photon decay or from the subsequent 
late time evolution as the energy propagates outward to infinity?  

In order to understand the above issues it was useful to study the propagation of radiation in strongly coupled holographic 
gauge theories.  A simple setting to study this problem is that of radiation produced 
by an accelerated quark in strongly coupled conformal field theories.  
If one takes a heavy quark and accelerates it along some given
trajectory, what does the pattern of emitted radiation look like?  
This problem is analogous to textbook studies of radiation produced by an accelerated 
charge in classical electrodynamics.
However, instead of radiation carried by a classical $U(1)$ gauge field, the radiation is that of a strongly coupled non-Abelian quantum field theory. 
Is the emitted radiation isotropic at strong coupling?  If not, does it isotropize as it propagates to infinity?  

This problem was studied in \cite{Athanasiou:2010pv,Hatta:2011gh}. There it was found that the radiation produced by the accelerated quark 
is not isotropic.   Nor does the radiation isotropize as it propagates out to infinity.  Remarkably, 
the angular distribution of power produced by an accelerated quark is the same at weak and strong coupling!
This suggests that the isotropy observed in the decay of off-shell photons at strong coupling comes from the 
production mechanism itself --- \textit{i.e.} the decay --- and not the subsequent evolution.  

The lesson from this is that one should not treat the jet production mechanism with strong coupling tools.  
Indeed, in asymptotically free QCD it is natural to expect the production mechanism during the early stages of 
heavy ion collisions to be governed by weakly coupled physics.
Ideally, one should treat the initial hard physics with perturbative QCD and then study soft processes 
such as energy loss with strongly coupled tools.
Simply put, the utility of modeling jets with strongly coupled tools comes from studying their evolution and energy loss
\textit{after} they are produced.  

\subsection{Holographic models of jets}

While at strong coupling it is impossible to create well collimated sprays of energy via off-shell photon decays, 
one can construct states at some initial time $t = t_i$ which share features with jets created via weakly coupled processes.
Namely, one can construct states with excitations which 
are well-collimated, have energy $E \gg T$ with $T$ the temperature of the plasma,
and which can propagate very far through the plasma before
thermalizing.  Ideally, one would compute the initial state using perturbative QCD.
However, we shall take a different approach and focus on 
universal features which are insensitive to the precise 
initial conditions used to construct the state at time $t_i$.  These include the rates of energy loss and momentum broadening
and the penetration depth defined as the total distance a jet can propagate through plasma before thermalizing.

Let us focus for simplicity on $q \bar q$ jets in strongly coupled conformal field theories (for other holographic models of jets see for example \cite{Chesler:2011nc,Arnold:2010ir,Arnold:2011qi,Arnold:2011vv}).  
According to holographic duality a $q \bar q$ pair moving through plasma is dual to a classical string moving through 
a black hole geometry \cite{Karch:2002sh}.  The geometry dual to an infinite static plasma at temperature $T$ is the 
AdS-Schwarzschild geometry, where the metric may be written 
\begin{equation}
\label{eq:adsbh}
ds^2 = \frac{L^2}{u^2} \left [ -f(u) dt^2 + d \bm x^2 + \frac{du^2}{f(u)} \right ],
\end{equation}
where $L$ is the AdS radius and $f(u) = 1 - u^4/u_h^4$.  The horizon of the geometry is at radial coordinate $u = u_h$ with $u_h = 1/\pi T$.  
The boundary of the geometry, which is where the dual field theory lives, is at radial coordinate $u = 0$.   

As it evolves in time
the string sweeps out and $1+1$ dimensional \textit{worldsheet}.  Events on the worldsheet can be labeled by coordinates $(\tau,\sigma)$
which are related to events in spacetime via the embedding functions $X^{M} \equiv \{t(\tau,\sigma),\bm x(\tau,\sigma),u(\tau,\sigma)\}$.
The dynamics of the embedding functions
are governed by the Nambu-Goto action
\begin{equation}
\label{eq:nabugoto}
S = - T_0 \int d \tau d \sigma \sqrt{-\gamma},
\end{equation}
where the string tension is $T_0 = \frac{\sqrt{\lambda}}{2 \pi L^2}$ with $\lambda$ the 't Hooft coupling and $\gamma = \det \gamma_{ab}$ with 
\begin{equation}
\label{eq:wsmetric}
\gamma_{ab} = \partial_a X \cdot \partial_b X,
\end{equation}
the worldsheet metric. Here and below the indices $(a,b)$ run over the worldsheet coordinates $\tau$ and $\sigma$.  
The action (\ref{eq:nabugoto}) is simply the area of the worldsheet.

Varying the action we obtain the string equations of motion 
\begin{equation}
\label{eq:stringeom}
\partial_\tau \Pi^{\tau}_\mu + \partial_\sigma \Pi^{\sigma}_\mu,
\end{equation}
where the string worldsheet currents $\Pi^{a}_M$ are
\begin{equation}
\label{eq:stringfluxes}
\Pi^a_M = \frac{\delta S}{\delta (\partial_a X^M)} = -T_0 \sqrt{-\gamma} \gamma^{ab} g_{MN} \partial_a X^N,
\end{equation}
where $g_{MN}$ is the AdS-Schwarzschild metric and as usual $\gamma^{ab}$ is the inverse of $\gamma_{ab}$.
Note that the energy of the string is 
\begin{equation}
\label{eq:stringenergy}
E = - \int d \sigma \, \Pi^\tau_0.
\end{equation}
The string equations of motion (\ref{eq:stringeom}) are simply the equations of energy 
conservation on the worldsheet with $\Pi^\sigma_0$ the energy flux down the string.

\begin{figure}
\begin{centering}
\includegraphics[width=7cm]{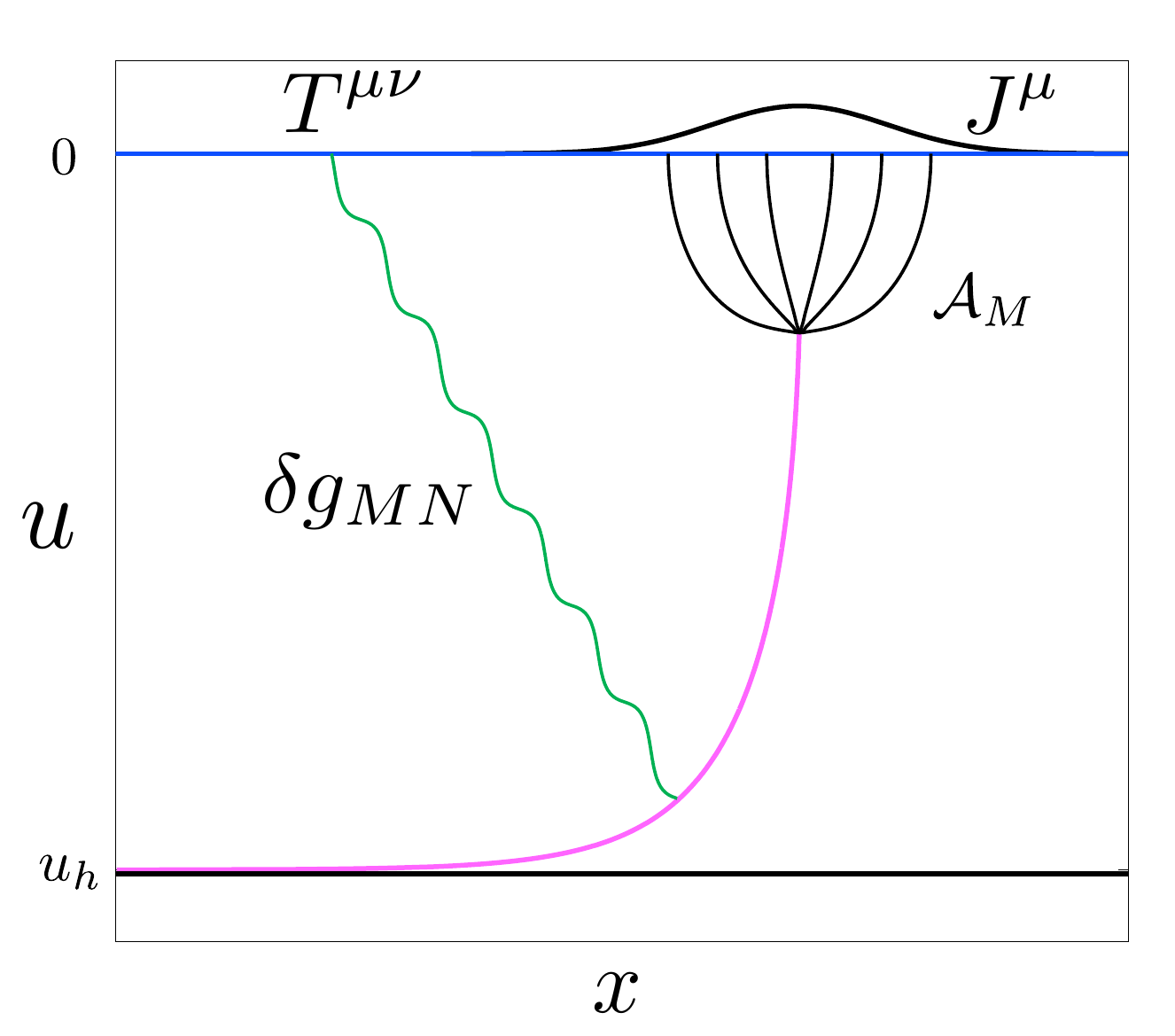}
\par\end{centering} 
\protect\caption{
\label{fig:stringcartoon} A cartoon showing a string at some fixed time in the 
AdS-Schwarzschild geometry.  The endpoint of the string is charged under a $U(1)$ gauge field 
$\mathcal A_M$.  The boundary of the AdS-Schwarzschild geometry --- located at radial coordinate $u = 0$ ---
behaves as an electromagnetic conductor.  Up to a minus sign the current $J^\mu$ induced on the boundary 
has the physical interpretation as the baryon current of the quark.  Likewise, via Einstein's equations 
the string perturbs the AdS-Schwarzschild geometry.  The near-boundary behavior of the metric perturbation
encodes the stress tensor $T^{\mu \nu}$ of the jet.}
\end{figure}

The endpoints of the string are charged under a bulk $U(1)$ gauge field $\mathcal A_M$.  The boundary 
of the AdS geometry behaves as an ideal conductor \cite{Chesler:2008wd}.  Hence the gauge field induces a current $J^\mu$ on the 
boundary, which up to an overall sign has the interpretation as the baryon current of a \textit{dressed} quark \cite{Chesler:2008wd,Chesler:2008uy}.  This is shown
schematically in Fig.~\ref{fig:stringcartoon}.  Therefore, one can regard the spatial location of the string endpoint as roughly coinciding with the spatial 
location of the quarks.  The boundary conditions at the string endpoints are set by the mass of the quark \cite{Karch:2002sh}.  
For infinitely massive quarks the endpoint 
is fixed at the boundary.  Correspondingly, the current $J^\mu$ is localized with delta function support.  In contrast, for massless quarks the endpoints are free to 
fall in the AdS-Schwarzschild geometry.  Likewise, the current $J^\mu$ spreads out as the endpoints fall.  For intermediate mass quarks the endpoint 
is fixed to be at some radial coordinate $u_m$.

Via Einstein's equations the presence of the string also perturbs the AdS-Schwarzschild geometry.  Just as the near boundary behavior of the gauge field $\mathcal A_M$ encodes
the quark baryon current $J^\mu$, the near-boundary behavior of the metric perturbation $\delta g_{MN}$ encodes the stress tensor $T^{\mu \nu}$
of the system.  This is also shown schematically in Fig.~\ref{fig:stringcartoon}.

With the above preliminaries layed out, the basic strategy then will be to construct initial string data at some time $t_i$  
and evolve the string profile forward in time according to the string equations of motion (\ref{eq:stringeom}).  To echo the previous point, 
we seek strings whose boundary boundary energy density $T^{0 0}$ is that of a well collimated spray of energy which propagates very far before thermalizing.
We will then study jet quenching and other observables which are insensitive to the details of the precise form of the initial data.
In what follows we will highlight results for infinitely massive quark jets and massless quark jets.

\subsection{Heavy quark energy loss}
One of the first studies \cite{Herzog:2006gh,Gubser:2006bz,CasalderreySolana:2006rq} of energy loss via AdS/CFT came from the study of heavy quarks propagating 
at constant velocity $v$ through a static box of plasma at temperature $T.$   As the quark moves through the plasma it transfers 
energy and momentum to the plasma via frictional drag.
Hence an external force is required to maintain constant velocity.  This can be supplied by an electromagnetic field coupled to 
the quark's baryon current $J^\mu$.

\begin{figure}[h!]
\begin{centering}
\includegraphics[width=13cm]{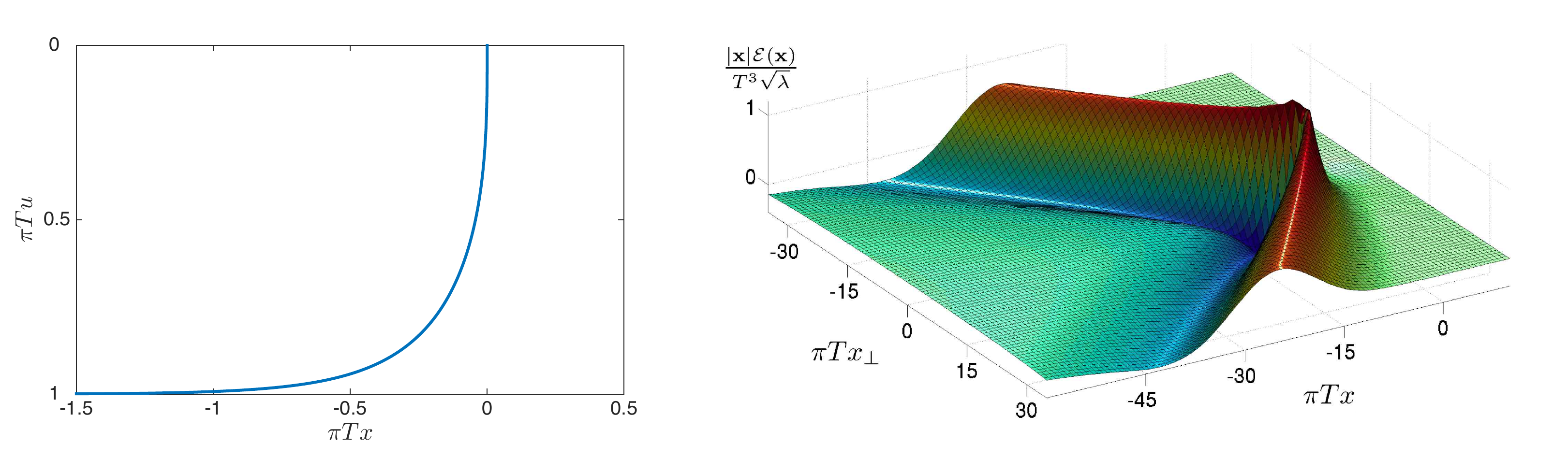}
\par\end{centering}
\protect\caption{\label{fig:massiveprofileandenergy} Left: the string profile (\ref{eq:trailingstring}) for a heavy quark.
Right: the energy density produced by a heavy quark moving at three quarters the speed of light in the $x-$direction.  A Mach cone is clearly present.}
\end{figure}

Under the assumption that the quark has been dragged at constant velocity for an arbitrary duration of time, it is natural 
to seek a steady-state solution to the string equations of motion.
Choosing worldsheet coordinates $\tau = t$ and $\sigma = u$ the steady-state ansatz take the form
\begin{equation}
\label{eq:heavyansatz}
x(t,u) = v t + v x_0(u),
\end{equation}
with the endpoint terminating at the boundary $u = 0$.  
Solving the string equations of motion (\ref{eq:stringeom}) for $x_0(u)$ yields 
\begin{equation}
\label{eq:trailingstring}
x_0(u) = \frac{1}{2 u_h} \left [ \tan^{-1} \frac{u}{u_h} + { \frac{1}{2}} \log \frac{u_h - u}{u_h + u} \right ].
\end{equation}
We plot $x_0(u)$ in Fig.~\ref{fig:massiveprofileandenergy}.

The electromagnetic field dragging the quark (and hence the string endpoint) must 
be supplying energy at a rate equal to the rate energy is dissipated by drag.  In the dual gravitational description
the energy lost via drag is encoded by a flux of energy down the string towards the event horizon, $\frac{dE}{dt} = \Pi^\sigma_0$.
Plugging the the solution (\ref{eq:trailingstring}) into Eq.~(\ref{eq:stringfluxes}), the drag reads \cite{Herzog:2006gh,Gubser:2006bz,CasalderreySolana:2006rq}
\begin{equation}
\label{eq:heavyloss}
\frac{dE}{dt} = { \frac{\pi}{2}} \sqrt{\lambda} T^2 \frac{v^2 }{\sqrt{1 - v^2}}.
\end{equation}
Hence, as the 't Hooft coupling $\lambda \to \infty$ the quark loses energy faster and faster.

A natural question to ask is where does the energy lost by the quark go?
To answer this question it is useful to study the stress tensor $T^{\mu \nu}$.  This requires solving Einstein's equations for the metric perturbation $\delta g_{MN}$ due 
to the presence of the string and extracting the stress tensor from its near-boundary asymptotics.  This calculation was carried out in \cite{Chesler:2007an,Chesler:2007sv,Gubser:2007ga,Friess:2006fk}.
In the right panel of Fig.~\ref{fig:massiveprofileandenergy} we also plot the normalized energy density $\mathcal E \equiv \frac{2 \pi^2}{N_c^2}T^{00}$ for a heavy quark moving in the $x-$direction at three quarters the speed of light.
The speed of sound in a relativistic conformal field theory is $c_s = 1/\sqrt{3} \approx 0.57.$  Hence the quark is moving supersonically.  As is clear from the figure,
the quark creates a Mach cone.  Evidently, the energy lost via drag is carried away by sound waves.

There have since been numerous studies of heavy quark energy loss at strong coupling in more general settings.  Some examples include 
heavy quark energy loss in non-conformal theories \cite{Ficnar:2011yj}, anisotropic plasmas \cite{Chernicoff:2012gu} and in far-from-equilibrium 
states such as the shock wave collisions discussed above \cite{Chesler:2013cqa}.  In the latter study it was demonstrated that even in 
far-from-equilibrium states, such as those produced by colliding shock waves, the equilibrium energy loss formula (\ref{eq:heavyloss}) provides a reasonable estimate of the heavy quark energy loss rate.
Additionally, heavy quark energy loss due to deceleration (\textit{i.e.} non steady-state) was studied in \cite{Chernicoff:2008sa},
and the rates of heavy quark momentum broadening were studied in \cite{CasalderreySolana:2007qw,Gubser:2006nz}.

\section{Light quark energy loss}

In contrast to strings dual to heavy quarks, where the endpoints are fixed to the AdS boundary,
the endpoints of strings dual to massless quarks are allowed to fall unimpeded towards the  
event horizon of the AdS-Schwarzschild geometry \cite{Karch:2002sh,Chesler:2008wd}.
For massless quarks the string equations of motion (\ref{eq:stringeom}) must be solved subject to the  
open string boundary conditions 
\begin{equation}
\label{eq:openbc}
\Pi^{\sigma}_M = 0 {\rm \ at \ the \ endpoints.}
\end{equation}

The fact that string endpoints fall toward the horizon implies there does not exist steady-state solutions to the string equations of motion.
Hence studying light quark energy loss is considerably more complicated than studying heavy quark energy loss.
Nevertheless it is a straightforward matter to numerically solve the string equations of motion (\ref{eq:stringeom}) with the open string boundary conditions (\ref{eq:openbc}).
For details on how to do this see for example \cite{Chesler:2008uy}.

\begin{figure}
\begin{centering}
\includegraphics[width=13cm]{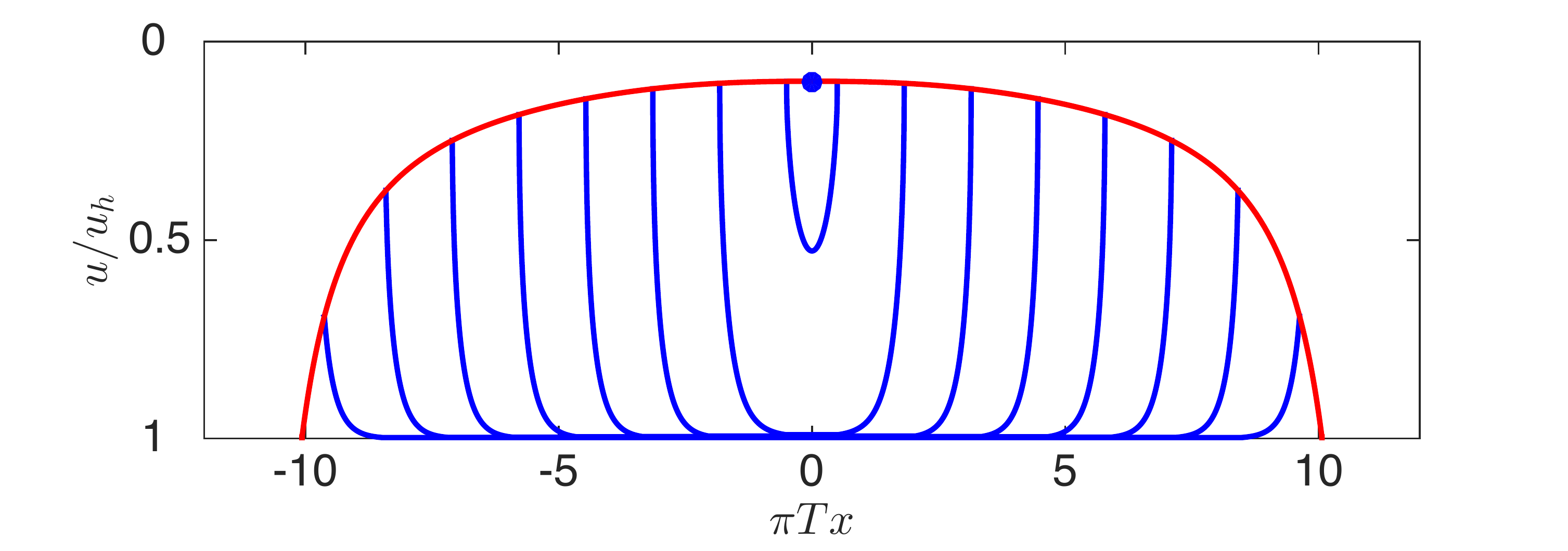}
\par\end{centering}

\protect\caption{A numerically generated falling string profile dual to a light quark jet.
The string is shown at several values of fixed coordinate time $t$.  
The string is created at the point $x = u = 0$ and subsequently expands into an extended object.
The red lines denote 
the string endpoint trajectories.  The endpoints travel a distance $x_{\rm stop} \approx 10/\pi T$
before falling into the horizon at $u = u_h$.
\label{fig:stringsnapshots}}
\end{figure}

In Fig.~\ref{fig:stringsnapshots} we plot a numerically generated string profile at several values of fixed coordinate time $t$.
At time $t = 0$ the initial string profile is simply a point located at the AdS boundary $u = 0$.
As time progresses the string expands from a point to an extended object and the endpoints 
fall towards the horizon.  The endpoints travel a total distance $x_{\rm stop} \approx 10/\pi T$ before falling
into the horizon.  As the endpoint fall into the horizon, the boundary baryon current $J^\mu$ (depicted schematically in Fig.~\ref{fig:stringcartoon}) 
becomes delocalized and starts to evolve hydrodynamical according to the diffusion equation \cite{Chesler:2008wd}.  
This corresponds to the thermalization of the light quark jet.

The total distance the string endpoints and hence the jet can travel depends on initial conditions.
A natural question is for a fixed energy $E$, what is the maximum distance the string endpoints can travel before falling into the horizon?
In the language of the dual field theory, for fixed jet energy $E$, what is the maximum distance a light quark jet can travel before thermalizing?
To answer this question one can numerically generate an ensemble of solutions to the string equations of motion (\ref{eq:stringeom}) with different initial conditions.
One can then compute the energy (\ref{eq:stringenergy})
and compare the stopping distance $x_{\rm stop}$ to $E$.  This computation was carried out in \cite{Chesler:2008uy} where it was found that the maximum distance 
a jet with energy $E$ can travel before thermalizing is given by 
\begin{equation}
\label{eq:stopingdistance}
x_{\rm stop} = \frac{\mathcal C}{T} \left (\frac{E}{T \sqrt{\lambda}} \right )^{1/3},
\end{equation}
where $C \approx 0.526$.  The $x_{\rm stop} \sim E^{1/3}$ scaling was also found in \cite{Gubser:2008as}.
The constant $\mathcal C$ was recently computed analytically and found to be  \cite{Ficnar:2013wba} 
\begin{equation}
\mathcal C = \frac{2^{1/3} \Gamma({\textstyle \frac{5}{4}})}{\sqrt{\pi}\Gamma({\textstyle \frac{3}{4}})}.
\end{equation}

We now turn to the light quark energy loss rate.  Following \cite{Chesler:2014jva}, we focus on strings whose $x_{\rm stop}$ is asymptotically	 
large compared to $1/T$.  Such string have worldsheets which are approximately null \cite{Chesler:2008uy,Chesler:2014jva}.  
Why?   When $x_{\rm stop} \to \infty$ the scaling (\ref{eq:stopingdistance}) requires $E\to \infty$.
Since strings have finite tension the $E \to \infty$ limit is generically realized by strings that expand at nearly the speed of light, 
meaning that the string profile must be approximately that of an expanding filament of
null dust.  Indeed, null strings have profiles $X_{\rm null}^M$ satisfying $\gamma(X_{\rm null}) = 0$
and from (\ref{eq:stringfluxes}) have divergent energy density.

Since null strings satisfy $\gamma(X_{\rm null}) = 0$ they minimize the Nambu-Goto action (\ref{eq:nabugoto}) 
and are exact albeit singular solutions to the string equations of motion (\ref{eq:stringeom}).  Following \cite{Chesler:2014jva}, to obtain finite energy
solutions to the equations of motion one can expand the string embedding functions about a null string solution
\begin{equation}
X^M = X^M_{\rm null} + \epsilon \delta X_{(1)}^M +  \epsilon^2 \delta X_{(2)}^M + \dots,
\end{equation}
where $\epsilon$ is a bookkeeping parameter (which can be set equal to 1 at the end of calculations).  In what follows it is useful to choose 
worldsheet coordinates $\tau = t$ and $\sigma$ such that $\dot X_{\rm null} \cdot X'_{\rm null} = 0$ and $\delta X_{(m)} = \{0,\delta \bm x,0 \}$.
We denote the location of the string endpoints by $\sigma = \sigma_*$.  The string equations of motion
can then be solved perturbatively in powers of $\epsilon$.

Focusing on strings propagating in the $x-$direction, the null string embedding functions can be written 
\begin{equation}
X_{\rm null}^M = \{t,x_{\rm geo}(t,\sigma),0,0,u_{\rm geo}(t,\sigma)\},
\end{equation}
where for each $\sigma$ $x_{\rm geo}$ and $u_{\rm geo}$ satisfy the null geodesic 
equations, which read
\begin{eqnarray}  
\label{eq:geo1}
\frac{\partial x_{\rm geo}}{\partial t} &=& \frac{f}{\xi}, \\
\frac{\partial u_{\rm geo}}{\partial t} &=& \frac{f\sqrt{\xi^2 -f }}{\xi},
\end{eqnarray} 
where $\xi = \xi(\sigma)$.  The parameter $\xi$ determines the initial inclination of the geodesics in the $x-u$ plane
and, more fundamentally, specifies the conserved spatial momentum associated with the geodesics, $f(u)^{-1} \partial x_{\rm geo}/\partial t = \xi^{-1}$.
The geodesic equations have the solution
\begin{equation}
\label{eq:geosol}
x_{\rm geo} = -\frac{u_h^2}{u_{\rm geo}} {_2}F_1\left ( {\textstyle \frac{1}{4},\frac{1}{2}; \frac{5}{4}; \frac{u_h^4}{\zeta u_{\rm geo}^4}} \right )
+ C(\sigma)
\end{equation}
where $_2 F_1$ is the Gauss hypergeometric function, $\zeta \equiv 1/(1 - \xi^2)$ and $C(\sigma)$ is an arbitrary function
which can be chosen such that $x_{\rm geo}(t = 0,\sigma) = 0$.

\begin{figure}
\begin{centering}
\includegraphics[width=12cm]{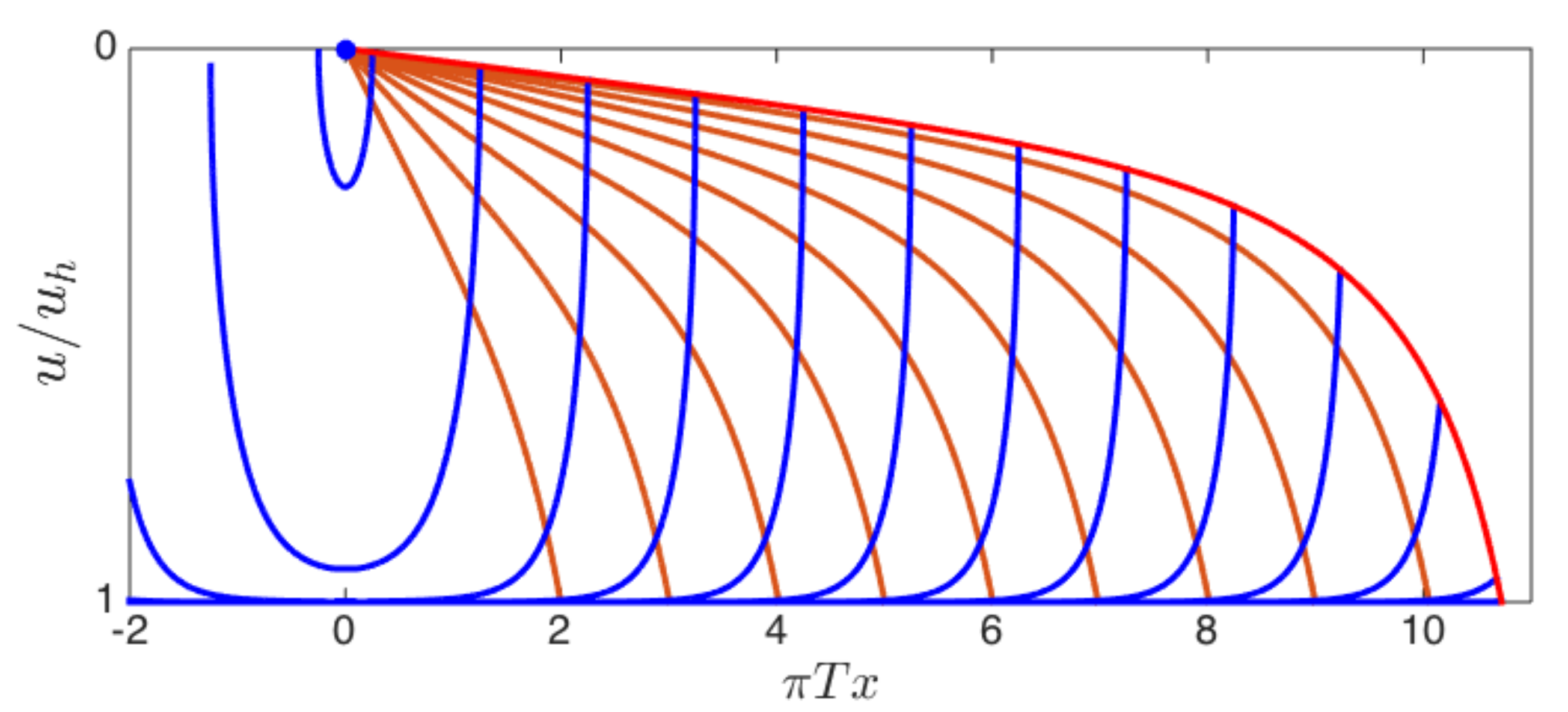}
\par\end{centering} 
\protect\caption{
\label{fig:NullString} A null string (blue) shown at several different coordinate times $t$.
The string starts off at a point on the boundary and expands at the speed of light while falling towards the horizon.
The rust colored curves represent the null geodesics that make up the string and the red curve is the endpoint trajectory.}
\end{figure}

Henceforth let us focus on strings created at the point $x = u = t = 0$ at the AdS boundary.  At asymptotically early times $t \ll u_h$, when the string is close to the AdS boundary,  $f = 1$ and  
the geodesics are given by 
\begin{eqnarray}
\label{eq:geosol2}
x_{\rm geo}  = t \cos \sigma, & u_{\rm geo} = t \sin \sigma.
\end{eqnarray}
Hence, the worldsheet coordinate $\sigma$ is simply the initial angle of the geodesic in the $x-u$ plane.  
Likewise,
\begin{equation}
\label{eq:xidef}
\xi(\sigma) = \sec \sigma.
\end{equation}

In Fig.~\ref{fig:NullString} we plot a null string generated by a congruence of 
geodesics with $\sigma_* = 0.025$.  The string profile, denoted by the blue curves, is shown at several values of coordinate time $t$.
The string starts off at a point on the boundary and expands at the speed of light while falling towards the horizon.
The rust colored curves represent the null geodesics that make up the string and the red curve is the endpoint trajectory.
The total distance the string endpoint can travel is entirely controlled by the angle $\sigma_*$.  As we elaborate on further below, 
in the limit $\sigma_* \to 0$ the stopping distance diverges.

To study energy loss rate we must compute the first order corrections to the string embedding functions.
At leading order in the bookkeeping 
parameter $\epsilon$ the worldsheet energy density $\Pi^\tau_0$ and flux $\Pi^\sigma_0$ read
\begin{eqnarray}
\label{eq:Pi00}
\Pi^\tau_{0} = - \frac{T_0 \xi \partial_\sigma u_{\rm geo}}{u_{\rm geo}^2} \sqrt{ \frac{- \xi}{2 \epsilon f \partial_t \delta x_{(1)}}} + O(\sqrt{\epsilon}), &
\Pi^\sigma_0 = O(\sqrt{\epsilon}).
\end{eqnarray}
Hence at leading order in $\epsilon$ the string equations of motion (\ref{eq:stringeom}) simply read $\partial_t \Pi^0_0 = 0$ so
$\Pi_0^\tau = \Pi^\tau_0(\sigma)$.
In other words, energy is simply transported on the congruence of geodesics which make up the null string!  

We define the energy 
\begin{equation}
\label{eq:totalE}
E(x) \equiv - \int_{\sigma_*}^{\sigma_H(x)} \Pi^0_0(\sigma),
\end{equation}
where $\sigma_H(x)$ is the $\sigma$ corresponding the the geodesic which impacts the horizon at $x$.
In particular, $\sigma_H(x = x_{\rm stop}) = \sigma_*$.
The energy $E(x)$ is simply the total string energy that makes it past the point $x$.  The energy loss rate
per unit length is
\begin{equation}
\label{eq:lightquarkenergyloss}
\frac{d E}{dx} = -\Pi^\tau_0(\sigma_H(x)) \sigma_H'(x).
\end{equation}
We discuss the precise boundary interpretation of $dE/dx$ below.
To compute $dE/dx$ we must compute $\Pi^\tau_0$ and $\sigma_H(x)$.

Because $\Pi^\tau_0$ is time independent it is only necessary to compute it at an initial time.
Near the AdS boundary, where $f = 1$ and the geodesics are given by (\ref{eq:geosol2}), the string equation of motion $\partial_t \Pi^\tau_0 = 0$ 
leads to the equation of motion for $\delta x_{(1)}$
\begin{equation}
\partial_t^2 \delta x_{(1)} + \frac{2}{t} \partial_t \delta x_{(1)} = 0,
\end{equation}
which has the solution 
\begin{equation}
\label{eq:stringsol2}
\delta x(t,\sigma) = \phi(\sigma) + \frac{1}{t} \psi(\sigma),
\end{equation}
for arbitrary functions $\phi(\sigma)$ and $\psi(\sigma)$.
The open string boundary conditions (\ref{eq:openbc}) 
require $\psi(\sigma_*) = 0$.

Substituting the string solution (\ref{eq:stringsol2})
and the geodesic solution (\ref{eq:geosol2}) and (\ref{eq:xidef}) into (\ref{eq:Pi00}) 
we obtain
\begin{equation}
\label{eq:pi2}
\Pi^\tau_0 = -T_0 \csc^2 \sigma \sqrt{\frac{\csc 2 \sigma \sin \sigma}{ \epsilon \psi(\sigma)}} + O(\sqrt{\epsilon}).
\end{equation}

With $\Pi^\tau_0$ computed and the congruence of null geodesics specified we can now compute $dE/dx$.  
But what is $\psi(\sigma)$!  The energy loss rate depends on an arbitrary function.  At first sight this seems like a disaster.  
Not only can we not describe the production mechanism of jets via holography but their energy loss rate, at least for massless jets, seems 
to be very sensitive to initial conditions.  However, this grim diagnosis turns out to be incorrect: the ambiguities in the light quark energy loss rate are largely transient 
effects which die out over time scales of order $1/T$.  These effects are negligible in the high energy limit when $x_{\rm stop} \to \infty$.

Let us consider $x_{\rm stop} \to \infty$ limit with $x/x_{\rm stop}$ fixed and first compute $\sigma_H(x)$.
$\sigma_H(x)$ can be computed from the geodesics solution (\ref{eq:geosol}) by setting $u_{\rm geo} = u_h$ 
and solving the resulting transcendental equation for $\sigma$.  With $\xi$ given by (\ref{eq:xidef}) the resulting 
equation can be solved analytically in the limit $\sigma \to 0$ (\textit{i.e.} for geodesics which travel very far before falling into the horizon).
The solution reads $\sigma_H(x) = \frac{u_h^2 \Gamma({\textstyle \frac{1}{4}})^2}{16 \pi x^2}.$
Likewise, setting $\sigma_H(x = x_{\rm stop}) = \sigma_*$, we see that the stopping distance is related to $\sigma_*$ via
\begin{equation}
\label{eq:xstop2}
x_{\rm stop} = \frac{u_h \Gamma({\textstyle \frac{1}{4}} )^2}{4 \sqrt{\pi \sigma_*}}.
\end{equation}
As previously advertised, $x_{\rm stop} \to \infty$ when $\sigma_* \to 0$.
We therefore may write
\begin{equation}
\label{eq:sigmaH}
\sigma_H(x) = \sigma_* \left ( \frac{x_{\rm stop}}{x} \right)^2.
\end{equation}
Therefore, when we take $x_{\rm stop} \to \infty$ with $x/x_{\rm stop}$ fixed, we are forced to take $\sigma_H \to 0$.
Simply put, when $x_{\rm stop} \to \infty$, energy is transported  along geodesics which originate 
asymptotically close to the string endpoint. 
Only these geodesics propagate to distances $x = O(x_{\rm stop})$. This behavior can clearly be seen in Fig.~\ref{fig:stringsnapshots}.

Therefore, to get the energy loss rate 
we may expand Eq.~(\ref{eq:lightquarkenergyloss}) about $\sigma_H \sim \sigma_* \to 0$.  The function $\psi(\sigma)$ may be approximated by 
$\psi(\sigma) \approx \psi'(\sigma_*) (\sigma -\sigma_*)$.  
Hence, near the string endpoint 
\begin{equation}
\label{eq:piapprox}
\Pi^\tau_0  \approx - \frac{T_0}{\sigma^2 \sqrt{2 \epsilon \psi'(\sigma_*)(\sigma - \sigma_*)}}.
\end{equation}
The energy loss will depend on the unknown parameter $\psi'(\sigma_*)$ which 
can be repackaged in terms of the initial energy 
\begin{equation}
\label{eq:initialE}
E(x = 0) = \frac{\pi T_0}{2 \sigma_*^{3/2} \sqrt{2 \epsilon \psi'(\sigma_*)}} + O(1/\sqrt{\sigma_*}).
\end{equation}
Substituting (\ref{eq:piapprox}) and (\ref{eq:xstop2}) into (\ref{eq:lightquarkenergyloss})
and dividing by (\ref{eq:initialE}) we then secure \cite{Chesler:2014jva}
\begin{equation}
\label{eq:lightquark}
\frac{1}{E(x = 0)} \frac{d E}{dx} = - \frac{4 x^2}{\pi x_{\rm stop}^2 \sqrt{x_{\rm stop}^2 - x^2}}.
\end{equation}
Eq.~(\ref{eq:lightquark}) together with the stopping distance (\ref{eq:stopingdistance}) provides a complete description of 
energy loss in terms of the initial energy, 't Hooft coupling $\lambda$ and temperature $T$.

We now return to the boundary interpretation of $dE/dx$.
It turns out that it is possible to solve the gravitational backreaction 
problem analytically in the long wavelength limit.%
\footnote
  {
  This will be expanded upon in a coming paper.
  }
In doing so the hydrodynamic limit of the stress tensor $T^{\mu \nu}_{\rm hydro}$ can be extracted
from the gravitational field perturbation in closed form.  How is the string energy loss rate (\ref{eq:lightquark}) encoded in $T^{\mu \nu}_{\rm hydro}$?
The answer is 
\begin{equation}
\partial_\mu T^{\mu \nu}_{\rm hydro} = f^\nu(x) \delta^3(\bm x - \hat x t),
\end{equation}
with $f^0 = f^x = \frac{d E}{dx} \theta(x_{\rm stop} - x)$.  Simply put, the energy loss rate (\ref{eq:lightquark})
describes the rate in which energy is transferred from the jet to hydrodynamic modes.

The light quark energy loss rate (\ref{eq:lightquark}) stands in stark contrast to the heavy quark counterpart (\ref{eq:heavyloss}).
Heavy quarks lose most of their energy in the early stages of their trajectory when they are moving fastest.  In contrast, 
from (\ref{eq:lightquark}) we see that the light quark energy loss rate actually increases as $x$ increases!   
This behavior, which was first pointed out in \cite{Chesler:2008uy}, is 
reminiscent of a Bragg peak where the energy loss rate near thermalization is highest.  Indeed, the energy loss 
rate (\ref{eq:lightquark}) diverges as $x \to x_{\rm stop}$.%
\footnote
  {
  The rate that energy flows into hydrodynamic modes is only defined over time and length scales
  $\gg 1/T$.  As such (\ref{eq:lightquark})
  should be regarded as the gradient expansion of the instantaneous jet energy loss rate.
  The gradient expansion of a function $f(x)$ can diverge even if $f(x)$ is itself finite.
  Hence, the instantaneous energy loss rate need not diverge as $x \to x_{\rm stop}$.
  }
  
Recently the light quark energy loss rate (\ref{eq:lightquark}) was employed in a phenomenological 
hybrid model of jet quenching which included both strong and weak coupling physics \cite{Casalderrey-Solana:2014bpa,Casalderrey-Solana:2014wca}.
In the hybrid model the jet consists of a shower of partons which can split as the jet evolves.  The splitting events were treated using perturbative QCD.  
Via soft interactions, in between 
splitings the partons can interact with the thermal medium and lose energy.  The energy loss rate of the individual partons was treated 
using the light quark energy loss rate (\ref{eq:lightquark}).  The coefficient of the $E^{1/3}$ scaling in the stopping distance (\ref{eq:stopingdistance}) was allowed to 
vary providing a one parameter fit to data.  This gave a good fit to the data provided that the resulting stopping distances are a factor 3 to 4 times larger in QCD than that in SYM at the same temperature and energy.  This is natural since SYM has more degrees of freedom for energetic partons to interact with than QCD.

\section{Discussion with future perspectives}

We hope to have shown some remarkable applications of holography to
heavy ion collisions, while emphasizing that these results were obtained
in a theory quite different from QCD itself. Most notably almost all
results are obtained in the limit of infinitely strong coupling. Nevertheless,
many of the works presented promise to have a degree of universality:
they may apply to a large range of strongly coupled gauge theories.
One particular example of such a hope is figure \ref{fig:rapidity},
which could display a universal rapidity profile of the plasma formed
in high energy collisions in a strongly coupled gauge theories. Indications
are that this profile can match experimental data for RHIC energies,
while being perhaps too narrow at LHC energies. 

The latter may also be expected; it was already known that LHC collisions
did not agree well with the holographic prediction of the total multiplicity
of charged particles \cite{Gubser:2008pc,Lin:2009pn}, as expanded
upon in subsection \ref{sub:Colliding-shock-waves}. One natural explanation
would be that at such high energies the coupling constant is intermediate,
thereby invalidating the assumption of an infinitely strong coupling.
Indeed models with weak coupling tend to give wider rapidity distributions,
so from RHIC to LHC we may be watching a cross-over. On the other
hand it is necessary to make the AdS/CFT models more realistic before
drawing a firm conclusion, which can for instance be done by having
less symmetry, preferably containing event-by-event fluctuations in
the initial profile.

While this review naturally restricts to heavy ion collisions, we
also wish to mention that holography is used for a much broader range
of strongly coupled systems. These include non-fermi liquids \cite{Faulkner:2009wj}
and holographic superconductors \cite{Hartnoll:2008vx}, whereby the
latter are believed to possibly have some relation with high temperature
superconductors.

Lastly, we wish to express our excitement for the coming time, with
several new experimental runs with different energies, different nuclei
and higher statistics. Including features such as event-by-event distributions,
non-trivial correlations, rare photon jets will give a very constraining
data set, which will make it possible to distinguish all scenarios
presented. This will hopefully give major lessons for both QCD, quark-gluon
plasma physics, and applications of the gauge/gravity duality in a
broad sense.

\section*{Acknowledgments}

We thank Krishna Rajagopal for interesting discussions and comments on the manuscript. WS wishes
to thank Michal Heller, David Mateos, Jorge Casalderrey, Paul Romatschke
and Scott Pratt for collaborations on much of the work presented here.
PC is supported
by the Fundamental Laws Initiative of the Center for the
Fundamental Laws of Nature at Harvard University.
WS is supported by the U.S. Department of Energy under grant Contract
Number DE-SC0011090.

\bibliographystyle{ws-rv-van}
\bibliography{references}

\printindex                         
\end{document}